\documentclass[aps,physrev,reprint,noeprint,superscriptaddress]{revtex4-2} 
\usepackage{bm, braket,amsmath,mathtools,comment}
\usepackage{graphicx,color}
\usepackage{hyperref}

\begin{document}
\title{Tensor-network Monte Carlo approach based on time-evolving block decimation}
\author{Shimpei Goto}
\email[]{shimpei.goto@phys.s.u-tokyo.ac.jp}
\affiliation{Department of Physics, The University of Tokyo, Tokyo 113--0033, Japan}
\author{Hidemaro Suwa}
\affiliation{Department of Physics, The University of Tokyo, Tokyo 113--0033, Japan}
\author{Synge Todo}
\affiliation{Department of Physics, The University of Tokyo, Tokyo 113--0033, Japan}
\affiliation{Institute for Physics of Intelligence, The University of Tokyo, Tokyo 113--0033, Japan}
\affiliation{Institute for Solid State Physics, The University of Tokyo, Kashiwa 227--8581, Japan}
\date{\today}
\begin{abstract}
    We propose a tensor-network Monte Carlo (TNMC) approach for unitary evolution following the compression sequence of the time-evolving block decimation (TEBD) algorithm.
    In the TNMC approach, the obtained results contain evaluable statistical errors rather than truncation errors, unlike ordinary singular-value-decomposition-based methods such as the TEBD algorithm.
    Consequently, one can estimate unbiased expectation values within statistical errors even with a finite bond dimension.
    Since the sampling scheme is introduced in the simulations of unitary evolution, the proposed Monte Carlo scheme may suffer from a sign problem.
    We observe that the sign problem can be mitigated by increasing the bond dimension.
    We apply the proposed TNMC approach to the Hamiltonian and the Floquet dynamics.
    Numerical experiments show that the TNMC approach can estimate accurate expectation values of observables even when the TEBD method with the same bond dimension cannot.
    The proposed approach can be a new direction for improving the classical simulatability of unitary evolution.
\end{abstract}
\maketitle
\section{Introduction\label{sec:introduction}}
The importance of numerically simulating unitary evolution of quantum systems is growing.
Time evolution from a slightly excited state has been an important tool to investigate the responses of quantum systems~\cite{white_real-time_2004,baez_dynamical_2020,gartner_time-dependent_2022}.
Quantum devices under development~\cite{blais_circuit_2021,koch_charge-insensitive_2007,schreier_suppressing_2008,kim_evidence_2023,bruzewicz_trapped-ion_2019,cirac_quantum_1995,sackett_experimental_2000,franke_quantum-enhanced_2023,saffman_quantum_2010,maller_rydberg-blockade_2015,manetsch_tweezer_2025} are designed to implement unitary evolution.
Ultracold atom experiments provide direct access to Hamiltonian dynamics~\cite{greiner_collapse_2002,trotzky_probing_2012,bernien_probing_2017,takasu_energy_2020-1,honda_observation_2025}.
To validate or propose these tools, devices, and experiments in general settings, numerical simulations of unitary evolution are essential methods.

In simulating unitary evolution, the first obstacle is the dimension of the Hilbert space that increases exponentially with the system size.
This obstacle can be evaded by using tensor-network approaches~\cite{schollwock_density-matrix_2011,cirac_matrix_2021,vidal_efficient_2003,vidal_efficient_2004,daley_time-dependent_2004,garcia-ripoll_time_2006,wall_out--equilibrium_2012,haegeman_unifying_2016,li_time-dependent_2024,zaletel_time-evolving_2015,czarnik_time_2018,czarnik_time_2019} that compress the exponentially large Hilbert space to a manageable dimension.
In this compression, the singular value decomposition (SVD) of a matrix is usually adopted, and the rank of the matrix, often called the bond dimension, is reduced to a manageable size by truncating singular modes.
The quality of the compression is strongly affected by the entanglement of a quantum state.
A representative example of such an approach is the time-evolving block decimation (TEBD) algorithm~\cite{vidal_efficient_2003,vidal_efficient_2004}, and results from these approaches contain truncation errors.
It is nontrivial to evaluate deviations induced by truncation errors in expectation values obtained by such methods.

In this paper, we propose a classical approach for simulating unitary evolution whose results contain numerically evaluable statistical errors, not truncation errors.
The starting point of our proposed approach is the tensor-network Monte Carlo (TNMC) approach~\cite{ferris_unbiased_2015,arai_all-mode_2023,todo_markov_2024} that introduces Monte Carlo sampling to evaluate the contraction of tensor networks~\footnote{The term ``tensor network Monte Carlo'' is used in different contexts such as the variational Monte Carlo evaluation of tensor-network contraction or the proposal of spin configurations from tensor-network contraction.}.
In the TNMC approach, one keeps all the singular vectors obtained during the SVD of matrices and stochastically selects the set of singular vectors to reduce the bond dimension.
By taking the sampling average for the selected singular vectors, the truncation errors vanish, and statistical errors emerge.
So far, the TNMC approach has been applied to the tensor renormalization group protocol~\cite{levin_tensor_2007,morita_calculation_2019} to evaluate the partition function of classical spin models.

We extend the TNMC approach to be applicable for evaluating the expectation values of unitarily evolved quantum states.
Since the TEBD algorithm is one of the most successful classical approaches for simulating unitary evolution, we use the algorithm as a guide in designing the compression sequence of the TNMC approach.
The proposed extended TNMC approach is applied to two typical unitary evolutions: the Hamiltonian and the Floquet dynamics.
For Hamiltonian dynamics, we adopt the Heisenberg chain.
We observe that the proposed TNMC approach can estimate expectation values accurately even when the TEBD results with the same bond dimension show visible discrepancies.
Since the Monte Carlo approach is applied to unitary evolution, a sign problem~\cite{pan_sign_2024} may occur in the proposed TNMC approach.
We also confirm that the sign problem can be mitigated by increasing the bond dimension.
For the Floquet dynamics, we simulate the kicked-Ising chain dynamics, including the dual-unitary case~\cite{bertini_exact_2019,bertini_entanglement_2019,piroli_exact_2020}.
In the dual-unitary case, entanglement emerges rapidly, and the TEBD algorithm shows poor performance.
In such cases, randomly generated projectors show much better performance compared to the TEBD-based projectors.
In the Floquet dynamics, we also observe that the TNMC approach reproduces accurate results even when the TEBD with the same bond dimension does not.

The rest of the paper is organized as follows. The proposed TNMC approach and how to construct TEBD-based projectors are given in Sec.~\ref{sec:methods}.
The results of the proposed TNMC method applied to the Hamiltonian and the Floquet dynamics are shown in Sec.~\ref{sec:results}.
In Sec.~\ref{sec:summary}, we summarize the obtained results and discuss possible improvements of the proposed approach.

\section{Tensor-network Monte Carlo approach for unitary evolution\label{sec:methods}}
\subsection{Projector-based formulation and Markov-chain Monte Carlo sampling}
\begin{figure}
    \includegraphics[width=0.95\linewidth]{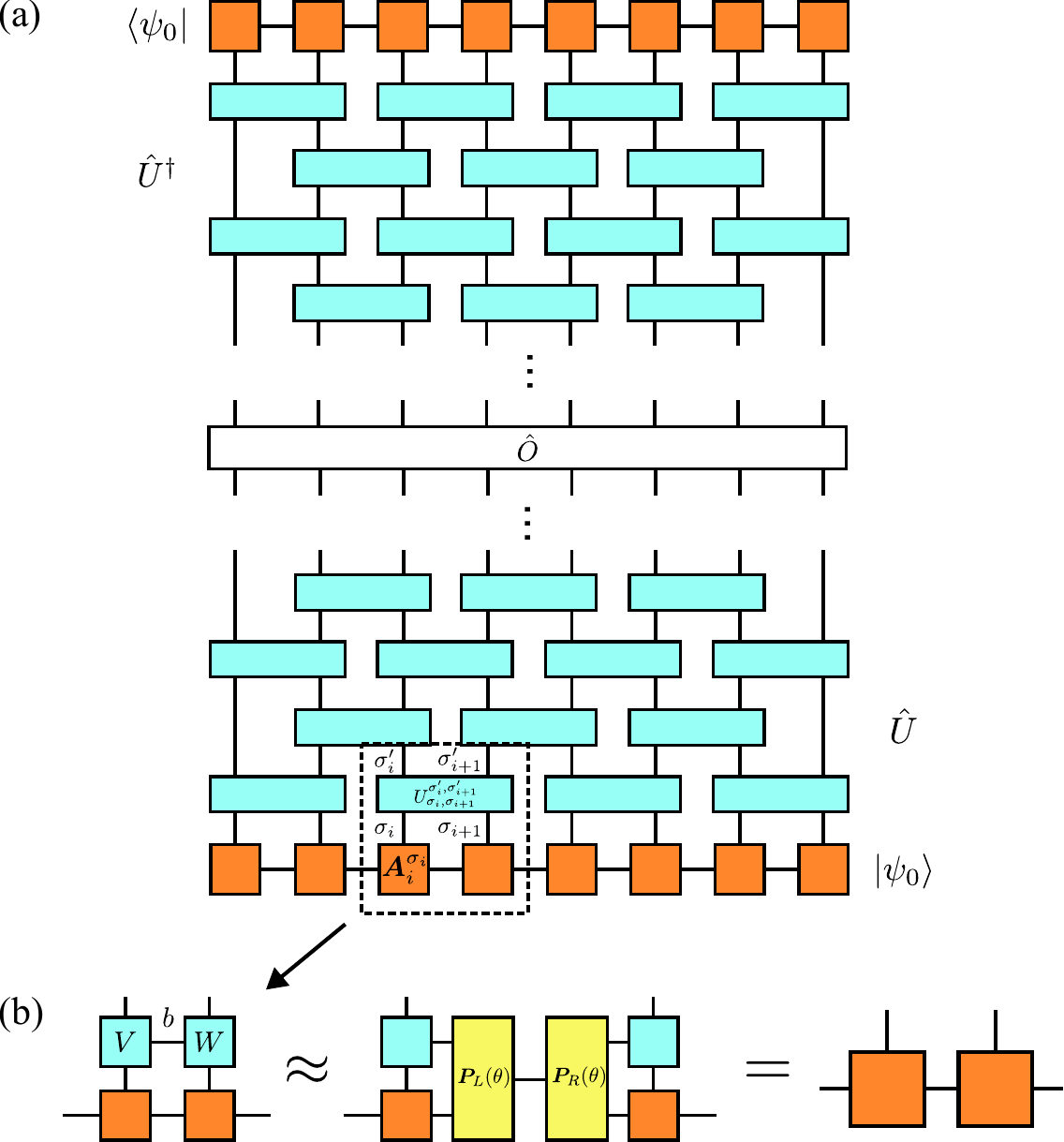}
    \caption{(a) Tensor network diagram to be evaluated via the TNMC approach in this study. The contraction of this tensor network gives the expectation value of an operator \(\hat{O}\) for a unitarily evolved state \(\hat{U}\ket{\psi_0}\), \(\braket{\psi_0|\hat{U}^\dagger \hat{O} \hat{U}|\psi_0}\). (b) Schematic diagram illustrating how an MPS is updated in the proposed TNMC approach.}
    \label{fig:network}
\end{figure}
The objective of the TNMC approach proposed in this study is to evaluate the contraction of the tensor network presented in Fig.~\ref{fig:network}, which gives the expectation value of an operator \(\hat{O}\) with a unitarily evolved state \(\hat{U}\ket{\psi_0}\), \(\braket{\psi_0|\hat{U}^\dagger \hat{O} \hat{U}|\psi_0}\).
Here, \(\ket{\psi_0}\) is an initial state and \(\hat{U}\) is a unitary operator that can be decomposed into neighboring unitary gates as depicted in Fig.~\ref{fig:network}.
Specifically, the unitary operator is given as
\begin{align}
    \hat{U} = \prod_{n=1} \hat{U}^\mathrm{even}_n \hat{U}^\mathrm{odd}_n,
\end{align}
where
\begin{align}
    \begin{aligned}
        \hat{U}^\mathrm{even}_n &= \prod_{i=1}\hat{U}(n)_{2i, 2i+1},\\
        \hat{U}^\mathrm{odd}_n &= \prod_{i=1}\hat{U}(n)_{2i-1, 2i}.
    \end{aligned}
\end{align}
Here, \(\hat{U}(n)_{i, i+1}\) is the \(n\)--th neighboring unitary gate acting on sites \(i\) and \(i+1\).
We assume that the matrix-product-state (MPS) representation~\cite{schollwock_density-matrix_2011} of the initial state \(\ket{\psi_0}\),
\begin{align}
    \ket{\psi_0} = \sum_{\bm{\sigma}} \bm{A}^{\sigma_1}_1 \bm{A}^{\sigma_2}_2 \ldots \bm{A}^{\sigma_N}_N\ket{\bm{\sigma}},
\end{align}
is available.
Here, \(\sigma_i\) represents the state of the local Hilbert space at the site \(i \), \(\ket{\bm{\sigma}} = \bigotimes^N_{i=1}\ket{\sigma_i}\), \(N\) is the number of sites, \(\sum_{\bm{\sigma}}\) means the summation over all possible configurations of \(\sigma_i\), and \(\bm{A}^{\sigma_i}_i\) denotes a \(\chi_{i-1} \times \chi_i\) matrix.
Hereafter, a bold capital letter represents a matrix.
The matrix dimension \(\chi_i\) is called the bond dimension, and the largest value among them is denoted by \(\chi \).
To make the product of matrices a scalar value, we set \(\chi_0 = \chi_N = 1\).
In such a situation, the TEBD~\cite{vidal_efficient_2003,vidal_efficient_2004} is one of the standard choices to evaluate the expectation value \(\braket{\psi_0|\hat{U}^\dagger \hat{O} \hat{U}|\psi_0}\).
Results obtained by the TEBD approach contain truncation errors coming from finite bond dimensions.
In contrast, the TNMC approach introduces statistical errors coming from the finite number of samples.

We formulate the TNMC approach for unitary evolution following the compression sequence of the well-tested TEBD algorithm.
After operating a neighboring unitary gate, a tensor is decomposed as
\begin{align}
    \begin{aligned}
    \bm{T}^{\sigma^\prime_i, \sigma^\prime_{i+1}} &= \sum_{\sigma_i, \sigma_{i+1}} U^{\sigma^\prime_i, \sigma^\prime_{i+1}}_{\sigma_i, \sigma_{i+1}}\bm{A}^{\sigma_i}_i \bm{A}^{\sigma_{i+1}}_{i+1} \\
    &= \bm{L}^{\sigma^\prime_i} \bm{R}^{\sigma^\prime_{i+1}}
    \end{aligned}
\end{align}
to keep the MPS representation.
The obtained matrix \(\bm{L}^{\sigma^\prime_i}\) (\(\bm{R}^{\sigma^\prime_{i+1}}\)) becomes the updated MPS representation \(\bm{A}^{\sigma_i}_i\) (\(\bm{A}^{\sigma_{i+1}}_{i+1}\)).
In the TEBD approach, this decomposition is performed by the  SVD of the tensor \(\bm{T}^{\sigma^\prime_i, \sigma^\prime_{i+1}}\), and singular modes with smaller singular values are truncated to keep the bond dimension manageable.

In the TNMC approach, the suppression of the bond dimension is carried out by applying a projector onto a stochastically chosen subspace.
To apply this projector, we first decompose the neighboring gate as
\begin{align}
    U^{\sigma^\prime_i, \sigma^\prime_{i+1}}_{\sigma_i, \sigma_{i+1}} = \sum_b V^{\sigma^\prime_i}_{\sigma_i, b} W^{\sigma^\prime_{i+1}}_{b, \sigma_{i+1}}
\end{align}
by the SVD without truncation.
The subscript \(b\) runs from 1 to \(r\), where \(r\) is the rank of the unitary gate reinterpreted as a matrix \(U_{(\sigma_i \sigma^\prime_i), (\sigma_{i+1} \sigma^\prime_{i+1})}\).
Then, we compose a projector onto a stochastically chosen  \(\tilde{\chi}_i\)-dimensional subspace \(\theta \) that can be represented as
\begin{align}
    \bm{P}^{b, b^\prime}(\theta) = \bm{P}^{b}_L(\theta)\bm{P}^{b^\prime}_R(\theta).
\end{align}
Here, \(\bm{P}^{b}_L(\theta)\) (\(\bm{P}^{b}_R(\theta)\)) is a \(\chi_i \times \tilde{\chi}_i\) (\(\tilde{\chi}_i \times \chi_i\)) matrix.
The dimension \(\tilde{\chi}_i \) becomes the bond dimension after the projection.
We also assume that the random average of projectors over stochastically chosen subspaces becomes the identity~\footnote{Although we call the tensor \(\bm{P}^{b, b^\prime}(\theta)\) a projector, this tensor does not need to be idempotent. The requirement for a ``projector'' in this paper is this identity condition.},
\begin{align}
    \begin{aligned}
    \Braket{\bm{P}^{b, b^\prime}(\theta)}_\theta &= \sum_\theta \bm{P}^{b, b^\prime}(\theta) p(\theta)\\
    &= \bm{I} \delta_{b, b^\prime}.
    \end{aligned}
    \label{eq:identity}
\end{align}
Here, \(p(\theta)\) is the probability to select a subspace \(\theta \), \(\bm{I}\) is the \(\chi_i \times \chi_i\) identity matrix, and \(\delta_{b, b^\prime}\) is the Kronecker delta.
The way of composing such a projector will be explained in the next sections.
With the composed projector, the decomposed matrices are given as
\begin{align}
    \begin{aligned}
        \bm{L}^{\sigma^\prime_i} &= \sum_{\sigma_i, b} V^{\sigma^\prime_i}_{\sigma_i, b} \bm{A}^{\sigma_i}_i \bm{P}^b_L(\theta),\\
        \bm{R}^{\sigma^\prime_{i+1}} &= \sum_{b, \sigma_{i+1}} W^{\sigma^\prime_{i+1}}_{b, \sigma_{i+1}} \bm{P}^b_R(\theta)\bm{A}^{\sigma_{i+1}}_{i+1}.
    \end{aligned}
    \label{eq:proj_update}
\end{align}
Repeating this projection, one can obtain the compressed MPS representation of a unitarily evolved unnormalized state \(\ket{\psi(\bm{\theta})}\), where \(\bm{\theta}\) represents the set of chosen subspaces during the projections.

Because of the identity condition~\eqref{eq:identity}, the random average of overlaps \(\braket{\psi(\bm{\phi})|\hat{O}|\psi(\bm{\theta})}\) gives the target quantity \(\braket{\psi_0|\hat{U}^\dagger\hat{O}\hat{U}|\psi_0}\).
We note that the bra state \(\ket{\psi(\bm{\phi})}\) and the ket state \(\ket{\psi(\bm{\theta})}\) should have independent projector sets for an unbiased estimation.
One can efficiently evaluate the overlaps since the unitarily evolved states are represented as MPSs.
A direct estimation of the random average of \(\braket{\psi(\bm{\phi})|\hat{O}|\psi(\bm{\theta})}\), however, can introduce large variance since the evolved states \(\ket{\psi(\bm{\theta})}\) and \(\ket{\psi(\bm{\phi})}\) are not normalized.

We circumvent this potential difficulty by introducing a weighted average implemented via a Markov-chain Monte Carlo approach.
The expectation value \(\braket{\psi_0|\hat{U}^\dagger\hat{O}\hat{U}|\psi_0}\) can be represented as the ratio of reweighted averages,
\begin{align}
\label{eq:observation}
\braket{\psi_0|\hat{U}^\dagger\hat{O}\hat{U}|\psi_0} &= \sum_{\bm{\phi}, \bm{\theta}} \braket{\psi(\bm{\phi})|\hat{O}|\psi(\bm{\theta})}q(\bm{\phi})q(\bm{\theta}) \nonumber \\
&= \frac{\sum_{\bm{\phi}, \bm{\theta}} \braket{\psi(\bm{\phi})|\hat{O}|\psi(\bm{\theta})}q(\bm{\phi})q(\bm{\theta})}{\sum_{\bm{\phi}, \bm{\theta}} \braket{\psi(\bm{\phi})|\psi(\bm{\theta})}q(\bm{\phi})q(\bm{\theta})} \nonumber \\
&= \frac{\frac{1}{Z}\sum_{\bm{\phi}, \bm{\theta}} \frac{\braket{\psi(\bm{\phi})|\hat{O}|\psi(\bm{\theta})}}{w(\bm{\phi}, \bm{\theta})}w(\bm{\phi}, \bm{\theta})q(\bm{\phi})q(\bm{\theta})}{\frac{1}{Z}\sum_{\bm{\phi}, \bm{\theta}} \frac{\braket{\psi(\bm{\phi})|\psi(\bm{\theta})}}{w(\bm{\phi}, \bm{\theta})}w(\bm{\phi}, \bm{\theta})q(\bm{\phi})q(\bm{\theta})}.
\end{align}
Here, \(q(\bm{\theta})\) is the probability to select the set of subspaces \(\bm{\theta}\), \(w(\bm{\phi}, \bm{\theta})\) is a positive weight function, and \(1/Z\) is the normalizing constant with \(Z = \sum_{\bm{\phi}, \bm{\theta}}w(\bm{\phi}, \bm{\theta})q(\bm{\phi})q(\bm{\theta})\).
The weighted average with the weight proportional to \(w(\bm{\phi}, \bm{\theta})q(\bm{\phi})q(\bm{\theta})\) can be estimated by the Metropolis-Hastings algorithm~\cite{metropolis_equation_1953,hastings_monte_1970}.
For the weight function, a candidate is
\begin{align}
    w(\bm{\phi}, \bm{\theta}) = \sqrt{\braket{\psi(\bm{\phi})|\psi(\bm{\phi})}\braket{\psi(\bm{\theta})|\psi(\bm{\theta})}}
    \label{eq:weight_overlap}
\end{align}
because the weighted observables are bounded as
\begin{align}
    \frac{\left|\braket{\psi(\bm{\phi})|\hat{O}|\psi(\bm{\theta})}\right|}{\sqrt{\braket{\psi(\bm{\phi})|\psi(\bm{\phi})}\braket{\psi(\bm{\theta})|\psi(\bm{\theta})}}} \leq \left|\lambda_\mathrm{max}\right|
\end{align}
from the Cauchy-Schwarz inequality.
Here, \(\lambda_\mathrm{max}\) is the largest absolute eigenvalue of \(\hat{O}\).
Consequently, the variance of the weighted observables is also bounded as
\begin{align}
    \mathrm{Var}\left[\frac{\mathrm{Re}\braket{\psi(\bm{\phi})|\hat{O}|\psi(\bm{\theta})}}{\sqrt{\braket{\psi(\bm{\phi})|\psi(\bm{\phi})}\braket{\psi(\bm{\theta})|\psi(\bm{\theta})}}}\right] \leq \left|\lambda_\mathrm{max}\right|^2.
    \label{eq:variance_bound}
\end{align}

The approach based on the weighted average breaks down when the denominator of the ratio~\eqref{eq:observation} is close to zero.
In this sense, the denominator plays the same role as an averaged sign in quantum Monte Carlo approaches~\cite{pan_sign_2024}.
A denominator that is too small indicates an insufficient bond dimension for ongoing simulations.

We note that the projective update~\eqref{eq:proj_update} does not depend on site indices other than \(i\) and \(i+1\).
When a projector between sites \(i\) and \(i+1\) is updated and the depth of unitary gates is \(\tau \), one has to repeat the update procedure~\eqref{eq:proj_update} \(\tau/2\) times on average.
Consequently, the complexity for updating all projectors scales as \(O(N \tau^2 \chi^3)\)\footnote{If one computes the overlap \(\braket{\psi(\bm{\theta})|\psi(\bm{\theta})}\) naively to evaluate the weight function, an additional factor \(N\) arises in the complexity. This additional factor can be eliminated by storing tensors appearing during the computation of the overlap as often performed in the computation of effective Hamiltonian in the density-matrix renormalization group approach~\cite{schollwock_density-matrix_2011}.}.

\subsection{Randomly generated projectors}
We first introduce a naive implementation for projectors: Randomly generated projectors.
In preparing a projector between sites \(i\) and \(i+1\), we generate an \(r \chi_i \times r \chi_i\) unitary matrix \(\bm{R}\) Haar randomly following the procedure given in Ref.~\cite{mezzadri_how_2007}.
From the random unitary matrix, matrices \(\bm{Q}^b_L \) \((1 \leq b \leq r)\) are obtained by reshaping the matrix \(\bm{R}\) into a \(\chi_i \times r \times r \chi_i\) tensor.
Similarly, matrices \(\bm{Q}^b_R\) are obtained by reshaping the matrix \(\bm{R}^\dagger \) into an \(r\chi_i \times r \times \chi_i\) tensor.

From the \(r\chi_i\) columns of \(\bm{Q}^b_L\), we choose \(\tilde{\chi}_i\) columns with equal probabilities for the compression.
To perform the compression, we construct an \(r\chi_i \times \tilde{\chi}_i\) selector matrix \(\bm{S}(\theta)\).
The subspace \(\theta \) can be specified by \(\tilde{\chi}_i\) column indices.
The non-zero entries of \(\bm{S}(\theta)\) are given by
\begin{align}
    S(\theta)_{\theta_n, n} = 1, \label{eq:selector}
\end{align}
where \(\theta_n\) is the index of the \(n\)-th selected column in the subspace \(\theta \).
By applying the selector tensor, matrices
\begin{align}
    \begin{aligned}
    \bm{K}^b_L(\theta) &= \bm{Q}^b_L \bm{S}(\theta),\\
    \bm{K}^b_R(\theta) &= \bm{S}(\theta)^T \bm{Q}^b_R
    \end{aligned}
\end{align}
are obtained, and a projector
\begin{align}
\bm{K}^{b, b^\prime}(\theta) = \bm{K}^b_L(\theta) \bm{K}^{b^\prime}_R(\theta)
\end{align}
can be constructed from the matrices.

Since the columns are selected with equal probabilities, the probability to select a subspace \(\theta \), \(p(\theta)\), is \(1 / \binom{r\chi_i}{\tilde{\chi}_i}\) for any \(\theta \).
In addition, the number of subspaces that contain a specific column \(k\) is \(\binom{r\chi_i - 1}{\tilde{\chi}_i -1}\) independent of \(k\).
From these counts and the unitarity of \(\bm{Q}^b_L\) and \(\bm{Q}^b_R\), the random average of the projector is given by
\begin{align}
    \sum_{\theta}\bm{K}^{b, b^\prime}(\theta) p(\theta) = \frac{\tilde{\chi}_i}{r\chi_i}\bm{I} \delta_{b, b^\prime}.
\end{align}
Consequently, we have to rescale the projector as
\begin{align}
    \begin{aligned}
    \bm{P}^b_L(\theta) &= \sqrt{\frac{r\chi_i}{\tilde{\chi}_i}}\bm{K}^b_L(\theta),\\
    \bm{P}^b_R(\theta) &= \sqrt{\frac{r\chi_i}{\tilde{\chi}_i}}\bm{K}^b_R(\theta)
    \end{aligned}
\end{align}
to satisfy the identity condition~\eqref{eq:identity}.
We note that the scaling factor \(r\chi_i / \tilde{\chi}_i\) is the inverse of the probability to select a certain row.
This is one simple instance of projectors that can be used for TNMC simulations.

\subsection{Time-evolving-block-decimation-based projectors}
The above randomly generated projectors do not use the information of dynamics.
In this section, we explain how to construct a projector following the TEBD algorithm.
In other words, we introduce a projector \(\bm{K}^{b, b^\prime}(\theta)\) that reproduces results obtained by the TEBD algorithm.

The starting point is the TEBD implementation proposed in Ref.~\cite{hastings_light-cone_2009}.
In this implementation, an initial MPS is transformed into the right-canonical form \(\bm{B}^{\sigma_{i}}_{i}\), and one stores a diagonal matrix \(\bm{\Lambda}_i\), whose entries are singular values obtained by the SVD of \(\bm{A}^{\sigma_{i+1}}_{i+1}\) during the right canonicalization.
In this TEBD update, one computes 
\begin{align}
    \bm{C}^{\sigma^\prime_i, \sigma^\prime_{i+1}} = \sum_{\sigma_i, \sigma_{i+1}} U^{\sigma^\prime_i, \sigma^\prime_{i+1}}_{\sigma_i, \sigma_{i+1}}\bm{B}^{\sigma_i}_i \bm{B}^{\sigma_{i+1}}_{i+1}
\end{align}
and
\begin{align}
    \bm{\Theta}^{\sigma^\prime_i, \sigma^\prime_{i+1}} = \bm{\Lambda}_{i-1} \bm{C}^{\sigma^\prime_i, \sigma^\prime_{i+1}}.
\end{align}
The standard TEBD algorithm computes only the matrix \(\bm{\Theta}^{\sigma^\prime_i, \sigma^\prime_{i+1}}\).
Next, one performs the (truncated) SVD of \(\bm{\Theta}^{\sigma^\prime_i, \sigma^\prime_{i+1}}\) like the standard TEBD as
\begin{align}
    \bm{\Theta}^{\sigma^\prime_i, \sigma^\prime_{i+1}} = \bm{X}^{\sigma^\prime_i} \tilde{\bm{\Lambda}}_i \bm{Y}^{\sigma^\prime_{i+1}\dagger}.
    \label{eq:svd_theta}
\end{align}
The matrix \(\bm{Y}^{\sigma^\prime_{i+1}\dagger}\) is right normalized even with truncation and thus can be treated as the updated \(\bm{B}^{\sigma_{i+1}}_{i+1}\).
The diagonal matrix \(\tilde{\bm{\Lambda}}_i\) is also the updated \(\bm{\Lambda}_i\).
To obtain the updated \(\bm{B}^{\sigma_i}_i\), one only has to compute
\begin{align}
    \sum_{\sigma^\prime_{i+1}} \bm{C}^{\sigma^\prime_i, \sigma^\prime_{i+1}} \bm{Y}^{\sigma^\prime_{i+1}} &= \sum_{\sigma^\prime_{i+1}} \bm{\Lambda}^{-1}_{i-1} \bm{\Theta}^{\sigma^\prime_i, \sigma^\prime_{i+1}}\bm{Y}^{\sigma^\prime_{i+1}}\nonumber \\
    &= \bm{\Lambda}^{-1}_{i-1} \bm{X}^{\sigma^\prime_i} \tilde{\bm{\Lambda}}_i
\end{align}
since this is the right canonical form of the remaining part of Eq.~\eqref{eq:svd_theta}, \(\bm{X}^{\sigma^\prime_{i}}\tilde{\bm{\Lambda}}_i\).
Consequently, one can update the operated MPS while keeping its right-canonical form.

Based on this TEBD implementation, we construct the matrices \(\bm{P}^b_L(\theta)\) and \(\bm{P}^b_R(\theta)\).
Initially, an initial MPS is transformed into right-canonical form, and \(\bm{\Lambda}_i\) is stored.
In applying a unitary gate, we compute the matrix \(\bm{\Theta}^{\sigma^\prime_i, \sigma^\prime_{i+1}}\) and perform the SVD of \(\bm{\Theta}^{\sigma^\prime_i, \sigma^\prime_{i+1}}\) without truncation.
Then, we compute the matrices \(\bm{Q}^b_L\) and \(\bm{Q}^b_R\) as
\begin{align}
    \bm{Q}^b_L = \sum_{\sigma_{i+1}, \sigma^\prime_{i+1}} W^{\sigma^\prime_{i+1}}_{b, \sigma_{i+1}} \bm{B}^{\sigma_{i+1}}_{i+1} \bm{Y}^{\sigma^{\prime}_{i+1}}
\end{align}
and 
\begin{align}
    \bm{Q}^b_R = \tilde{\bm{\Lambda}}^{-1}_{i}\sum_{\sigma_{i}, \sigma^\prime_{i}}\bm{X}^{\sigma_i \dagger} V^{\sigma^\prime_{i}}_{\sigma_{i}, b} \bm{\Lambda}_{i-1}\bm{B}^{\sigma_{i}}_{i} .
    \label{eq:Q_R}
\end{align}
One can easily check that
\begin{align}
    \sum_{\sigma_i, b} V^{\sigma^\prime_i}_{\sigma_i, b} \bm{B}^{\sigma_i}_i \bm{Q}^b_L = \bm{\Lambda}^{-1}_{i-1} \bm{X}^{\sigma^\prime_i} \tilde{\bm{\Lambda}}_i
\end{align}
and
\begin{align}
    \sum_{b, \sigma_{i+1}} \bm{Q}^b_R W^{\sigma^\prime_{i+1}}_{b, \sigma_{i+1}} \bm{B}^{\sigma_{i+1}}_{i+1} = \bm{Y}^{\sigma^\prime_{i+1}\dagger}.
\end{align}
Thus, the application of these matrices reproduces the TEBD result without truncation.

The truncation can be performed by applying a \(D \times \tilde{\chi}_i \) selector matrix \(\bm{S}(\theta)\)~\eqref{eq:selector} as well as the randomly generated projectors.
Here, \(D\) is the number of columns of \(\bm{Q}^b_L\), and the subspace \(\theta \) is specified by \(\tilde{\chi}_i \) singular mode indices of \(\bm{\Theta}^{\sigma^\prime_i, \sigma^\prime_{i+1}}\) in this case.
With the selector matrix, we construct matrices
\begin{align}
    \bm{K}^b_L(\theta) = \bm{Q}^b_L \bm{S}(\theta)
    \label{eq:P_L}
\end{align}
and
\begin{align}
    \bm{K}^b_R(\theta) = \bm{S}^\top(\theta) \bm{Q}^b_R.
    \label{eq:P_R}
\end{align}
When the singular modes with the largest singular values are chosen, the projected MPS representation
\begin{align}
    \begin{aligned}
    \tilde{\bm{B}}^{\sigma_i}_i &= \sum_{\sigma^\prime_i, b} V^{\sigma_i}_{\sigma^\prime_i, b} \bm{B}^{\sigma^\prime_i}_i \bm{K}^b_L(\theta),\\
    \tilde{\bm{B}}^{\sigma_{i+1}}_{i+1} &= \sum_{b, \sigma^\prime_{i+1}} W^{\sigma_{i+1}}_{b, \sigma^\prime_{i+1}} \bm{K}^b_R(\theta) \bm{B}^{\sigma^\prime_{i+1}}_{i+1}
    \end{aligned}
\end{align}
is the same as the updated MPS obtained by the TEBD algorithm.
From these matrices, we obtain a projector \(\bm{K}^{b, b^\prime}(\theta) = \bm{K}^b_L(\theta) \bm{K}^{b^\prime}_R(\theta)\).
Since \(\tilde{\bm{B}}^{\sigma_i}_i\) keeps the right-canonical form, the above procedure can be repeated with the projected MPS representation.

A singular mode is stochastically selected with a probability proportional to \(v^c \) following the procedure presented in Ref.~\cite{todo_markov_2024}.
Here, \(v\) is a corresponding singular value.
An exponent \(c\) is set to unity in this study.
Since the probability depends on a singular mode, the identity condition~\eqref{eq:identity} does not hold with the projector \(\bm{K}^{b, b^\prime}(\theta)\).
To make the projector fulfill the identity condition, we introduce a weighted selector matrix \(\bm{S}_w(\theta)\),  whose non-zero entries are given by
\begin{align}
    S_w(\theta)_{\theta_n, n} = \frac{1}{r(\theta_n)}.
\end{align}
Here, \(r(\theta_n)\) is the probability to select a singular mode \(\theta_n\)~\cite{todo_markov_2024}.
With the weighted selector, we introduce a weighted projector \(\bm{P}^{b, b^\prime}(\theta) = \bm{P}^{b}_L(\theta) \bm{P}^{b^\prime}_R(\theta)\), where
\begin{align}
    \begin{aligned}
        \bm{P}^b_L(\theta) &= \bm{Q}^b_L \bm{S}_w(\theta),\\
        \bm{P}^b_R(\theta) &= \bm{K}^b_R(\theta).
    \end{aligned}
\end{align}
This weighted projector satisfies the identity condition~\eqref{eq:identity}.

Because of the weighting \(1/r(\theta_n)\), the overlap \(\braket{\psi(\bm{\theta})|\psi(\bm{\theta})}\) becomes extremely large when modes with low probabilities are selected multiple times within one trial.
Such a large overlap can induce a severe autocorrelation problem when the weight function \(w(\bm{\theta}, \bm{\phi})\) is given as Eq.~\eqref{eq:weight_overlap}.
To relax the autocorrelation, we add a decay factor to the weight function as
\begin{align}
    w(\bm{\phi}, \bm{\theta}) &= \sqrt{\braket{\psi(\bm{\phi})|\psi(\bm{\phi})}\braket{\psi(\bm{\theta})|\psi(\bm{\theta})}}\nonumber\\
    &\times e^{-\gamma\frac{\braket{\psi(\bm{\phi})|\psi(\bm{\phi})}\braket{\psi(\bm{\theta})|\psi(\bm{\theta})}}{\braket{\psi(\bm{\theta}_0)|\psi(\bm{\theta}_0)}^2}}
    \label{eq:weight_decay}
\end{align}
with a small decay constant \(\gamma \).
Here, \(\bm{\theta}_0\) denotes the set of subspaces with the largest singular values.
The introduction of the decay factor removes the bound for the variance~\eqref{eq:variance_bound}, and a decay constant that is too large may induce a large variance.

\subsection{Rank deficiency in preparing projectors}
To fulfill the identity condition~\eqref{eq:identity}, \(\bm{Q}^{b}_L\) and \(\bm{Q}^{b}_R\) should be of full rank as  \(r \chi_{i} \times r \chi_{i}\) matrices, and thus \(D\) should be \(r \chi_{i}\).
This condition is always satisfied with the randomly generated projectors.
For the TEBD-based projectors, however, \(D\) can be smaller than \(r \chi_{i}\) and thus \(\bm{Q}^{b}_L\) and \(\bm{Q}^{b}_R\) become rank deficient.
In such cases, one has to fill deficient linear spaces by adding randomly generated vectors orthonormal to already obtained singular modes.
The orthonormal random vectors can be obtained by orthonormalization algorithms such as the modified Gram-Schmidt procedure~\cite{golub_matrix_2012}.
One also has to add a tentative singular value for an added random mode to determine the probability to select the random mode. 
For these randomly generated modes, the weighting \(1/r(\theta_i)\) is equally distributed to the matrices \(\bm{P}^b_L(\theta)\) and \(\bm{P}^{b^\prime}_R(\theta)\).

One possibility of rank deficiency is the computation of the (pseudo-) inverse \(\tilde{\bm{\Lambda}}^{-1}_{i}\) in Eq.~\eqref{eq:Q_R}.
One has to ignore too small singular values to make the computation of the inverse numerically stable.
Furthermore, too small singular values also lead to too small probability to select corresponding singular modes.
Another possibility is that the number of singular modes is insufficient from the beginning.
When the dimensions of the local Hilbert spaces for the sites \(i\) and \(i+1\) are \(d\), the rank of the unitary gate \(r\) is \(d^2\) in general.
In the SVD given in Eq.~\eqref{eq:svd_theta}, the rank is at most \(\min(d\chi_{i-1}, d\chi_{i+1})\).
Consequently, more than \(d^2 \chi_i - \min(d\chi_{i-1}, d\chi_{i+1})\) random modes should be added to fulfill the identity condition~\eqref{eq:identity}.
In this study, we keep singular values larger than \(v_\mathrm{max} r_\mathrm{inv}\) for computing the inverse \(\tilde{\bm{\Lambda}}^{-1}_{i}\).
Here, \(v_\mathrm{max}\) is the largest singular value obtained in the SVD of \(\bm{\Theta}^{\sigma^\prime_i, \sigma^\prime_{i+1}}\).
A singular value for random modes \(v_\mathrm{rnd}\) is determined so that the probability to select a random mode \(r(\theta_\mathrm{rnd})\) becomes a predetermined value \(p_\mathrm{rnd}\).
To determine the singular value \(v_\mathrm{rnd}\), we apply the bisection method to the logarithm of \(v_\mathrm{rnd}\).
If the determined singular value is larger than the smallest kept singular value, a singular value for random modes is set to half of the smallest kept singular value. 

Since the addition of random modes introduces ambiguity such as tentative singular values, we want to reduce the number of random modes.
For \(d = 2\) cases, the Kraus-Cirac decomposition of the two-site unitary gate exists~\cite{kraus_optimal_2001}, i.e., a unitary gate acting on sites \(i\) and \(j\), \(\hat{U}_{i, j}\), can be decomposed as
\begin{align}
    \hat{U}_{i, j} = \hat{M}_i \hat{M}_j\hat{R}^{XX}_{i, j}(\theta_X)\hat{R}^{YY}_{i,j}(\theta_Y)\hat{R}^{ZZ}_{i,j}(\theta_Z)\hat{N}_i \hat{N}_j.
\end{align}
Here, \(\hat{M}_i\) and \(\hat{N}_i\) are unitary operators acting on the site \(i\), and \(\hat{R}^{XX}_{i,j}(\theta_X)\) is a two-site XX rotation operator given by
\begin{align}
    \hat{R}^{XX}_{i, j}(\theta_X) &= e^{i \theta_X \hat{X}_i \hat{X}_j} \nonumber \\ 
    &= \cos\theta_X \hat{I}_i \hat{I}_j + i \sin\theta_X \hat{X}_i \hat{X}_j.
\end{align}
Here, \(\hat{I}_i\) is the identity operator acting on the site \(i\) and \(\hat{X}_i\) is the Pauli-X operator acting on the site \(i\).
The other two-site Pauli rotations \(\hat{R}^{YY}_{i,j}(\theta_Y)\) and \(\hat{R}^{ZZ}_{i,j}(\theta_Z)\) are also given similarly,
\begin{align}
    \hat{R}^{YY}_{i,j}(\theta_Y) &= \cos\theta_Y \hat{I}_i \hat{I}_j + i \sin\theta_Y \hat{Y}_i \hat{Y}_j,\\
    \hat{R}^{ZZ}_{i,j}(\theta_Z) &= \cos\theta_Z \hat{I}_i \hat{I}_j + i \sin \theta_Z \hat{Z}_i \hat{Z}_j.
\end{align}
Here, \(\hat{Y}_i\) (\(\hat{Z}_i\)) is the Pauli-Y (Pauli-Z) operator acting on the site \(i\).
These two-site Pauli rotations have matrix-product operator form with bond dimension 2, e.g.,
\begin{align}
    \hat{R}^{XX}_{i,j}(\theta_X) =
    \begin{pmatrix}
        \cos\theta_X \hat{I}_i & i\sin\theta_X \hat{X}_i
    \end{pmatrix}
    \begin{pmatrix}
        \hat{I}_j \\ \hat{X}_j
    \end{pmatrix}.
\end{align}
Consequently, the rank of the two-site Pauli rotation \(r\) is only \(d\), not \(d^2\).
For \(d=2\) cases, one can decompose a unitary gate into three rank-2 unitary gates and significantly reduce the number of random modes in preparing the TEBD-based projector.

\section{Results of tensor-network Monte Carlo approaches\label{sec:results}}
\subsection{Hamiltonian dynamics of the Heisenberg chain}
We first apply the above TNMC approach to the Hamiltonian dynamics, where the TEBD works well at least for a short time.
The Hamiltonian we adopt is the Heisenberg chain,
\begin{align}
    \hat{H} = J \sum_{i=1}\left( \hat{X}_i\hat{X}_{i+1} + \hat{Y}_i\hat{Y}_{i+1} + \hat{Z}_i \hat{Z}_{i+1}\right).
    \label{eq:Heisenberg}
\end{align}
Here, \(J\) is the exchange energy between two-level systems.
We divide the Hamiltonian~\eqref{eq:Heisenberg} into odd and even parts as
\begin{align}
    \begin{aligned}
    \hat{H}_\mathrm{odd} &= J \sum_{i=1}\left( \hat{X}_{2i-1}\hat{X}_{2i} + \hat{Y}_{2i-1}\hat{Y}_{2i} + \hat{Z}_{2i-1} \hat{Z}_{2i}\right), \\
    \hat{H}_\mathrm{even} &= J \sum_{i=1}\left( \hat{X}_{2i}\hat{X}_{2i+1} + \hat{Y}_{2i}\hat{Y}_{2i+1} + \hat{Z}_{2i} \hat{Z}_{2i+1}\right),
    \end{aligned}
\end{align}
and obtain Trotter gates,
\begin{align}
    \begin{aligned}
    \hat{T}^\mathrm{odd}_{\Delta t} &= e^{-i\Delta t \hat{H}_\mathrm{odd}/\hbar}\\
    &= \prod_{i=1}\hat{R}^{XX}_{2i-1, 2i}(\varphi_t)\hat{R}^{YY}_{2i-1, 2i}(\varphi_t)\hat{R}^{ZZ}_{2i-1, 2i}(\varphi_t),\\
    \hat{T}^\mathrm{even}_{\Delta t} &= e^{-i\Delta t \hat{H}_\mathrm{even}/\hbar}\\
    &= \prod_{i=1}\hat{R}^{XX}_{2i, 2i+1}(\varphi_t)\hat{R}^{YY}_{2i, 2i+1}(\varphi_t)\hat{R}^{ZZ}_{2i, 2i+1}(\varphi_t).
    \end{aligned}\label{eq:trotter_gates}
\end{align}
Here, \(\Delta t\) denotes the discretized time step and \(\varphi_t = -J\Delta t / \hbar \).
The discretized step \(\Delta t\) is set to \(0.025\hbar / J\) in this study.
With the Trotter gates, the unitary operator \(\hat{U}\) is composed following the second-order Suzuki-Trotter decomposition~\cite{hatano_finding_2005} as
\begin{align}
    \hat{U} = \hat{T}^\mathrm{odd}_{\Delta t/2}\left(\hat{T}^\mathrm{even}_{\Delta t}\hat{T}^\mathrm{odd}_{\Delta t} \right)^{N_t-1}\hat{T}^\mathrm{even}_{\Delta t}\hat{T}^\mathrm{odd}_{\Delta t/2},
\end{align}
where \(N_t\) is the total number of time steps.
As shown in Eq.~\eqref{eq:trotter_gates}, each Trotter gate is decomposed into three two-site Pauli rotations.
Consequently, the total depth of this unitary operator \(\tau \) is \(6 N_t + 3\).
An initial state is set to the N\'eel product state
\begin{align}
    \ket{\psi(0)} = \ket{0}\ket{1}\ket{0}\ket{1}\ldots,
\end{align}
where \(\ket{0}\) (\(\ket{1}\)) denotes the ground (excited) state of a local two-level system.
To avoid the underestimation of statistical errors coming from the autocorrelation~\cite{sokal_monte_1997}, we use the blocking analysis~\cite{gubernatis_quantum_2016} with a block size of 1024.
The system size \(N\) is set to 20.

\begin{figure}
    \includegraphics[width=1.0\linewidth]{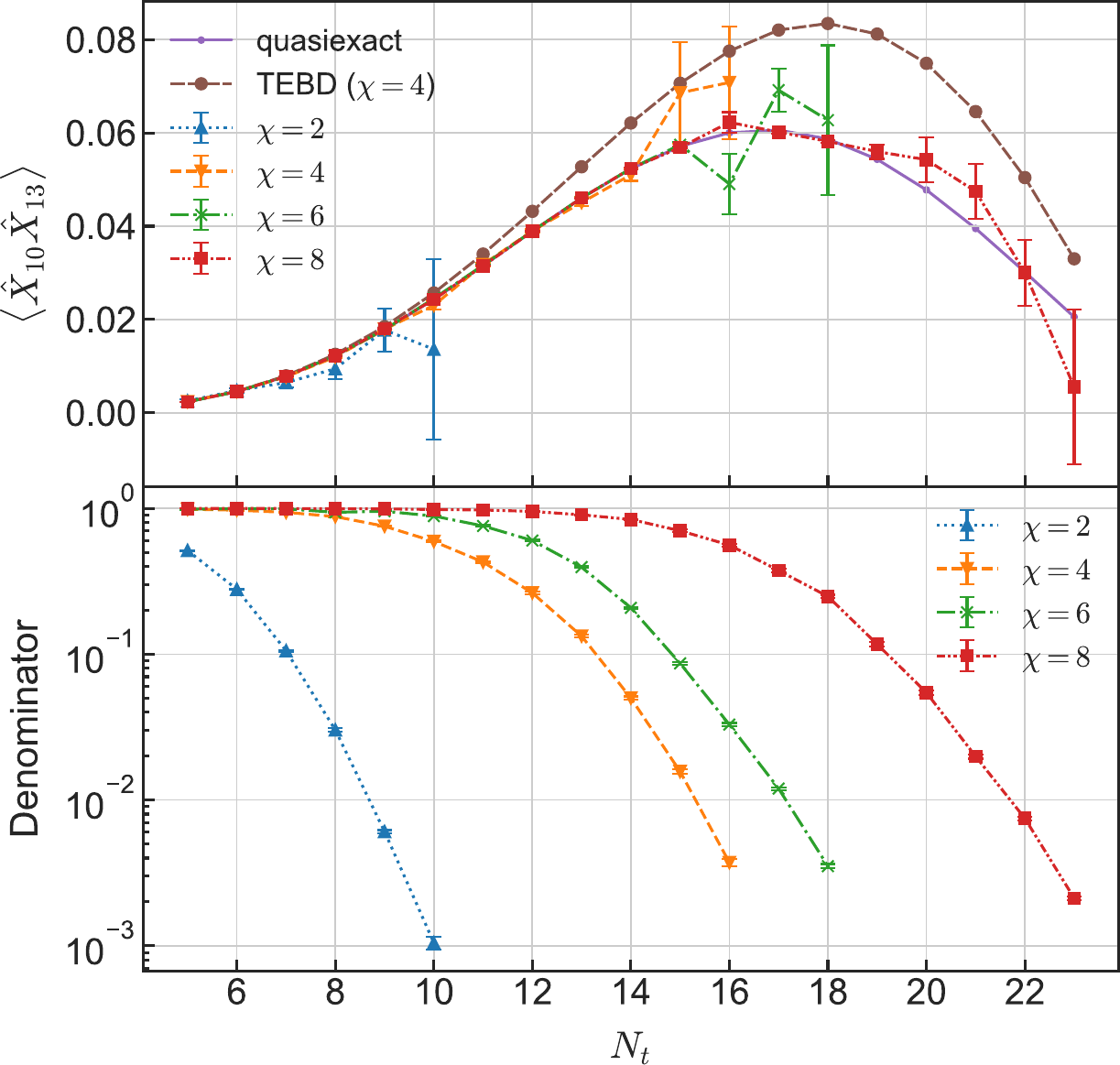}
    \caption{(Upper panel) Expectation value \(\braket{\hat{X}_{10} \hat{X}_{13}}\) and (lower panel) the denominator obtained by the TNMC approach with several bond dimensions as functions of the number of steps \(N_t\). The error bars indicate \(1\sigma\) uncertainty. The number of samples is 262144. We also plot the TEBD results with a bond dimension of \(\chi = 4\) and the quasiexact results. The quasiexact results are obtained by the TEBD method with an SVD threshold of \(10^{-30}\).\label{fig:tebd}}
\end{figure}

Figure~\ref{fig:tebd} shows the dependence of the expectation value \(\braket{\hat{X}_{10} \hat{X}_{13}}\) on the number of steps \(N_t\) obtained by the TNMC approach with several bond dimensions.
The expectation values and the statistical errors are estimated from 262144 samples.
The first thousand samples are discarded.
The parameters used for preparing projectors are summarized in Table~\ref{table:params}.
The decay constant \(\gamma \) of Eq.~\eqref{eq:weight_decay} is set to \(10^{-10}\) for all cases.
The observed lowest acceptance ratio of the Metropolis-Hastings algorithm is 0.988.
Except for two cases (\(N_t = 6\) and 8 with \(\chi =6\)), the estimated integrated autocorrelation time is less than 500.
With the setting, one can observe that the TNMC approaches reproduce the quasiexact results within the statistical errors.
Especially for the region \(12 \leq N_t \leq 14\) in the \(\chi = 4\) case, the TNMC approach gives accurate estimations even though the TEBD results with the same bond dimension show visible discrepancies~\footnote{In the TEBD simulations, the Kraus-Cirac decomposition is not performed.}.
The results obtained by the presented TNMC approach contain measurable statistical errors rather than truncation errors, as intended.

\begin{table}
    \caption{\label{table:params} Parameters \(r_{\mathrm{inv}}\) and \(p_{\mathrm{rnd}}\) used for simulations with bond dimension \(\chi \) presented in Fig.~\ref{fig:tebd}. The parameter \(r_{\mathrm{inv}}\) is used to determine threshold singular values for computing pseudo-inverse matrices. The parameter \(p_{\mathrm{rnd}}\) is a target probability for selecting a random mode to determine tentative singular values for random modes.}
    \begin{ruledtabular}
    \begin{tabular}{lll}
        \(\chi \) & \(r_{\mathrm{inv}}\) & \(p_{\mathrm{rnd}}\) \\
        \hline
        2 & \(10^{-6}\) & \(10^{-3}\) \\
        4 & \(10^{-8}\) & \(10^{-5}\) \\
        6 & \(10^{-8}\) & \(10^{-5}\) \\
        8 & \(10^{-8}\) & \(10^{-6}\)
    \end{tabular}
    \end{ruledtabular}
\end{table}

The statistical errors in Fig.~\ref{fig:tebd} increase as the number of steps \(N_t\) increases.
One can observe the exponential decrease of the denominator with the number of steps in Fig.~\ref{fig:tebd}.
In short, the increase in statistical errors is a consequence of the sign problem.
To keep the scale of statistical errors, the required number of samples increases exponentially with the number of steps.
Even though the exponential decrease of the denominator is inevitable, the decrease can be postponed by increasing the bond dimension.
In other words, the sign problem can be mitigated by increasing the bond dimension.

\subsection{Floquet dynamics of the kicked Ising chain}
\begin{figure}
    \includegraphics[width=0.6\linewidth]{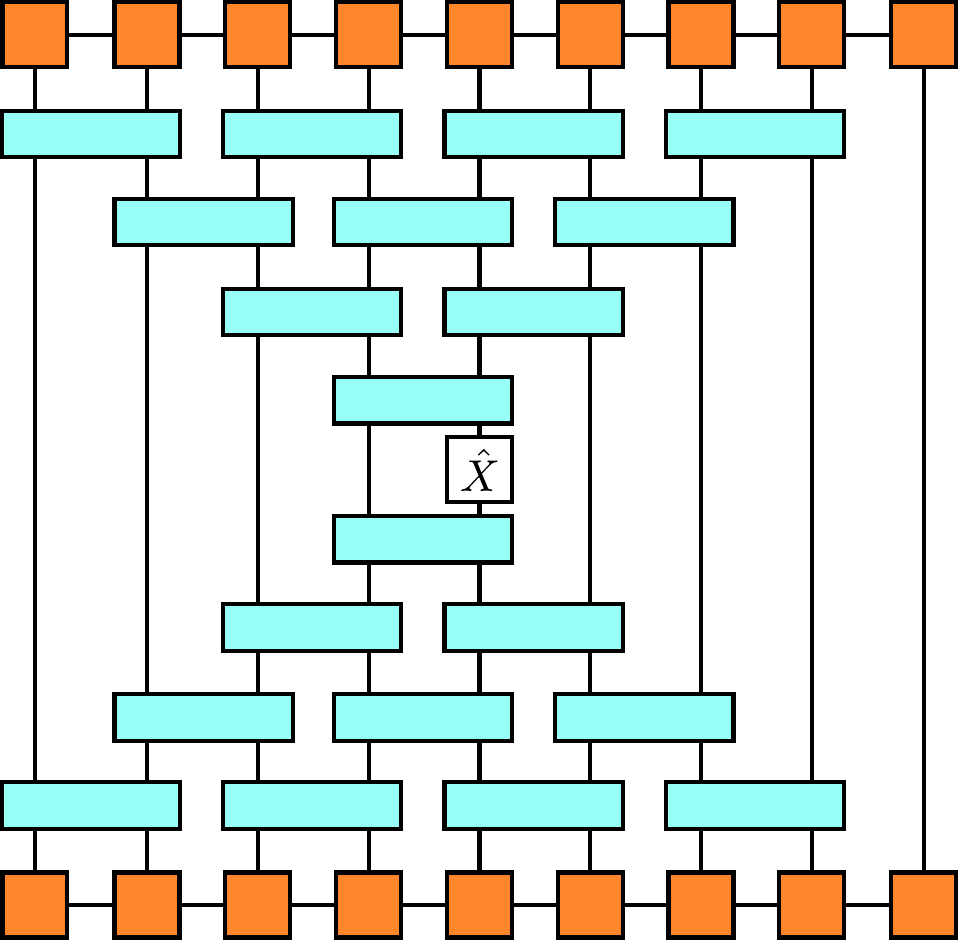}
    \caption{Lightcone structure used in the simulations of the Floquet dynamics of the kicked Ising chain.\label{fig:lightcone}}
\end{figure}

Next, we apply the TNMC approach to the Floquet dynamics of the self-dual kicked Ising chain~\cite{bertini_exact_2019,piroli_exact_2020,bertini_entanglement_2019}, where the TEBD shows poor performance because of the rapid increase of entanglement.
Although this Floquet dynamics is chaotic and nonintegrable, the analytic expressions of some correlation functions are available due to the dual-unitary character.
The neighboring unitary gates for the Floquet dynamics of the kicked Ising chain are given as~\cite{fischer_dynamical_2026-1}
\begin{align}
    &\hat{U}(n)_{i, i+1} \nonumber \\
    &\quad = e^{-ih\hat{Z}_i}e^{-iJ_z\hat{Z}_i\hat{Z}_{i+1}}e^{-ib(\hat{X}_i + \hat{X}_{i+1})}e^{-iJ_z\hat{Z}_i\hat{Z}_{i+1}}e^{-ih\hat{Z}_i}.
\end{align}
Here, \(h\), \(J_z\), and \(b\) are parameters that characterize the Floquet dynamics.
For \(|J_z| = |b| = \pi/4\), this unitary operator becomes dual unitary irrespective of \(h\)~\cite{bertini_entanglement_2019}.
The system size \(N\) is assumed to be odd, and an initial state is set to
\begin{align}
    \ket{\psi(0)} = \frac{1}{\sqrt{2}}(\ket{0} + \ket{1}) \bigotimes^{(N-1)/2}_{i=1}\ket{\mathrm{Bell}_{2i, 2i+1}},
\end{align}
because the analytical expressions of some correlation functions are available with this initial state.
Here, \(\ket{\mathrm{Bell}_{2i, 2i+1}}\) represents the Bell state
\begin{align}
    \ket{\mathrm{Bell}_{2i, 2i+1}} = \frac{1}{\sqrt{2}}(\ket{0}\ket{0} + \ket{1}\ket{1})
\end{align}
formed between sites \(2i\) and \(2i+1\).
The expectation value of the operator \(\hat{X}_i\) is given as
\begin{align}
    \braket{\hat{X}_i} = 
    \begin{cases}
        [\cos(2h)]^\tau & \text{if \(i = \tau+1 \)}\\
        0 & \text{otherwise}
    \end{cases}
\end{align}
when the unitary operator is dual unitary and the circuit depth is \(\tau \)~\cite{fischer_dynamical_2026-1}.
Since we only evaluate the local expectation value \(\braket{\hat{X}_{\tau+1}}\), the circuit is reduced to a lightcone structure as depicted in Fig.~\ref{fig:lightcone}.
The system size is set to 21. This value is irrelevant as long as it is larger than the size of the lightcone \(2 \tau \).

We apply the TNMC with the TEBD-based projector to the dual unitary Floquet dynamics (\(J_z = b = \pi/4\)) with the depth of unitary gates \(\tau = 5\).
In preparing the projectors, we set the parameters \(r_\mathrm{inv}\) and \(p_\mathrm{rnd}\) to \(10^{-6}\) and \(0.1\), respectively.
Since the observed autocorrelation is negligible, the decay factor and the blocking analysis are not used in following simulations.
The maximum bond dimension is set to 14.
The other parameter \(h\) is set to \(0.1 \pi \).
Thus, the expectation value \(\braket{\hat{X}_{\tau+1}}\) should be \(\left[ \cos(0.2\pi)\right]^5 \simeq 0.3466\).
With 1048576 samples, the TNMC approach estimates the expectation value \(\braket{\hat{X}_{\tau+1}}\) as \(0.2597 \pm 0.1718\).
Even though the exact value is contained within 1\(\sigma \) uncertainty, the statistical error seems too large for practical estimation.
This large error stems once again from the sign problem.
The estimated value of the denominator is only \(1.34\times 10^{-4} \pm 2.5 \times 10^{-5}\).
The denominator is too small.
We observe that the smaller parameter \(p_\mathrm{rnd}\) leads to the smaller denominator.

The TNMC approach with the TEBD-based projectors performs poorly in cases where standard TEBD fails.
In such cases, we find that a randomly generated projector shows much better performance.
This behavior can be understood by the singular value distribution of the MPS.
At the dual unitary point, the distribution of the singular values is almost flat.
Therefore, a completely random choice would be more suitable for the compression of the MPS than choosing the largest singular modes following the TEBD.

\begin{figure}
    \includegraphics[width=0.85\linewidth]{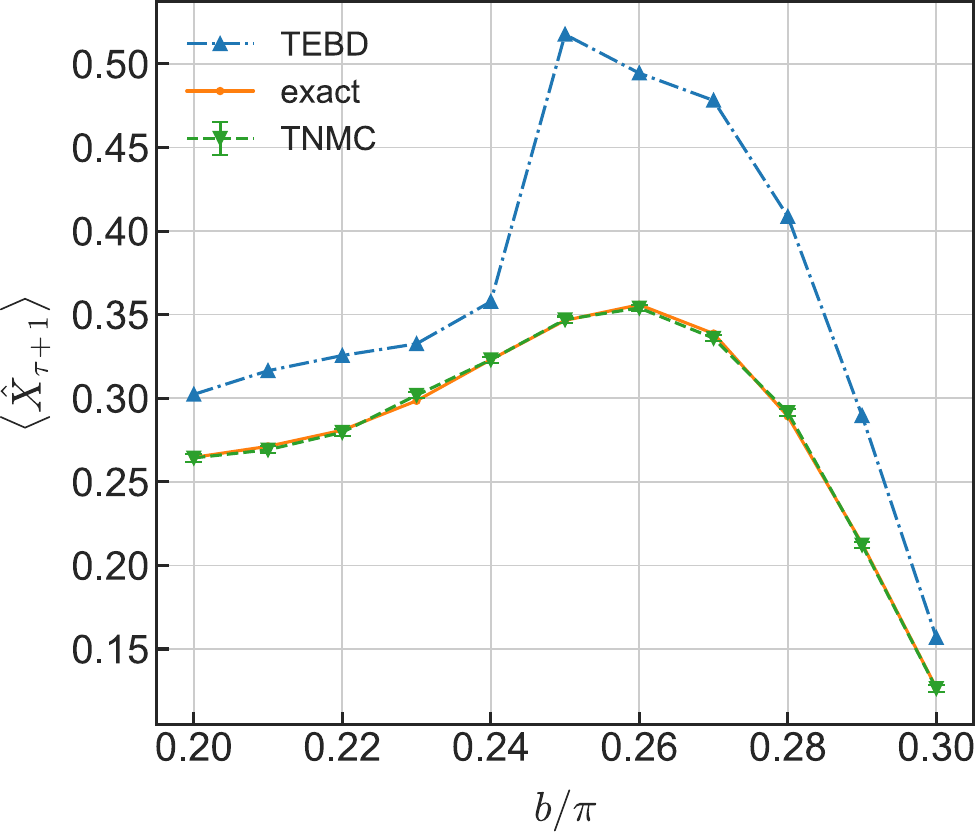}
    \caption{Expectation value \(\braket{\hat{X}_{\tau+1}}\) for depth \(\tau = 5\) as a function of the parameter \(b\). The parameters \(J_z\) and \(h\) are set to \(\pi/4\) and \(0.1\pi \), respectively. The maximum bond dimension is set to 14. The number of samples is 65536. The error bars denote \(1\sigma \) uncertainty. The TEBD result is obtained with the same maximum bond dimension. The exact results are obtained by the TEBD method with a maximum bond dimension of \(\chi = 32\).\label{fig:du_random}} 
\end{figure}

Figure~\ref{fig:du_random} shows the parameter \(b\) dependence of the expectation value \(\braket{\hat{X}_{\tau+1}}\) obtained by the TNMC with the random projectors.
The TNMC with the random projectors accurately reproduces the exact values.
We do not perform the sample average over randomly generated matrices \(\bm{Q}^b_L\) and \(\bm{Q}^b_R\).
For the other parameters of the simulations, we use the same values as those of the above TNMC with the TEBD-based projectors.
By changing the projector, the denominator estimated from 65536 samples at the dual-unitary point is increased to \(3.6769 \times 10^{-2} \pm 8.4 \times 10^{-5}\).
The estimated expectation value \(\braket{\hat{X}_{\tau+1}}\) is \(0.3471 \pm 0.0018\): the statistical error becomes around 1/100 with only 1/16 of the number of samples.
We also show the TEBD results with the same maximum bond dimension that exhibit large deviations. 
In such a situation where the TEBD shows poor performance, randomly generated projectors show much better performance compared to TEBD-based projectors.
Consequently, an adequate way for preparing projectors depends on the nature of unitary dynamics one tries to simulate.

\begin{figure}
    \includegraphics[width=0.7\linewidth]{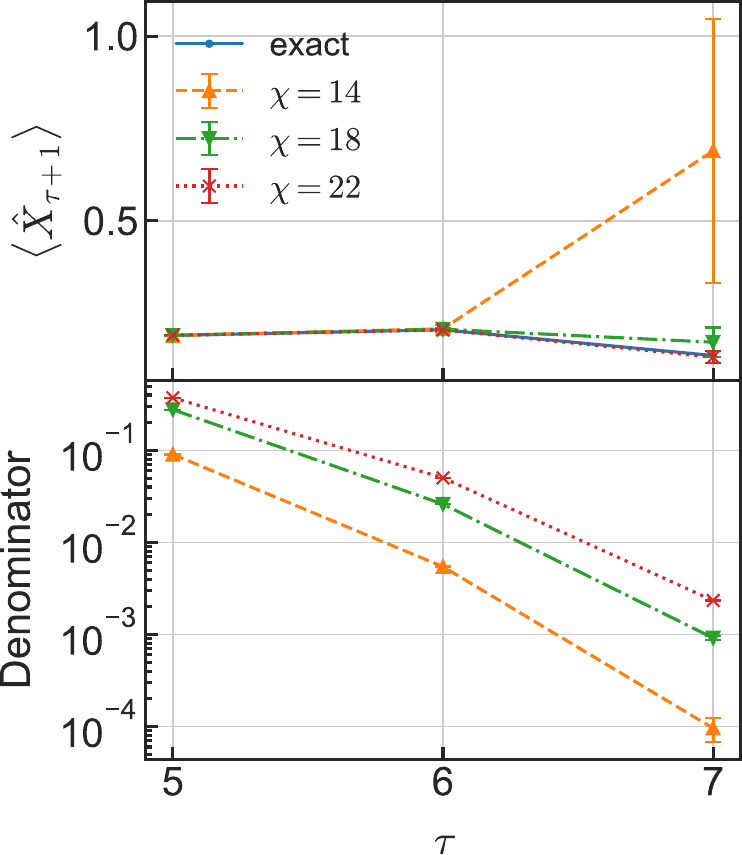}
    \caption{(Upper panel) Expectation value \(\braket{\hat{X}_{\tau+1}}\) and (lower panel) the denominator obtained by the TNMC approach with several bond dimensions as functions of the depth \(\tau \). The parameter \(b\) is set to \(0.15 \pi \). The number of samples is 262144. The other parameters used for simulations are the same as those in Fig.~\ref{fig:du_random}. The exact results are obtained by the TEBD method with a maximum bond dimension of \(\chi = 2^\tau \).\label{fig:du_tau_dep}}
\end{figure}

We also perform simulations with larger depths.
Figure~\ref{fig:du_tau_dep} represents the depth \(\tau \) dependence of the expectation value \(\braket{\hat{X}_{\tau+1}}\) and the denominator.
The parameter \(b\) is set to 0.15 \(\pi \).
For the other gate parameters, the same values as the above simulations are used.
Although the TNMC approach gives correct estimations for the \(\chi = 14\) case, the denominator decreases exponentially, and thus the statistical error increases exponentially with the depth.
The denominator increases significantly by slightly increasing bond dimension.
Like the Hamiltonian dynamics case, the sign problem can be mitigated by increasing bond dimension even in the strongly entangling Floquet dynamics.

\section{Summary and outlook\label{sec:summary}}
We proposed a tensor-network Monte Carlo (TNMC) approach for unitary evolution following the compression sequence of the time-evolving block decimation (TEBD) algorithm.
For projectors used in the proposed TNMC approach, we considered two types: the TEBD-based and the randomly generated projectors.
We applied the proposed TNMC approach to the Hamiltonian and the Floquet dynamics.
For the Hamiltonian dynamics of the Heisenberg chain, the TEBD algorithm performs well at least for a short time.
Reflecting this fact, the TNMC approach with the TEBD-based projectors showed good performance and reproduced quasiexact results.
We also observed that the TEBD results with the same bond dimension exhibited visible discrepancies.
For the Floquet dynamics, we chose the dynamics induced by the kicked-Ising chain, including the dual-unitary point.
Near the dual-unitary point, the entanglement evolves rapidly, and the TEBD algorithm performs poorly.
Consequently, the TNMC approach with the TEBD-based projector also showed poor performance.
To improve performance, we tried randomly generated projectors.
As a result, the TNMC approach with the randomly generated projectors showed much better performance and reproduced exact results even at the dual-unitary point.

The development of the TNMC approach for unitary evolution is still at an early stage, and there is much room for improving the efficiency of the approach.
For instance, the complexity of the present implementation has quadratic scaling with respect to the circuit depth \(\tau \).
This quadratic scaling emerges because we judge whether we accept or reject an updated projector once a projector and a matrix-product state are updated.
Since the update scheme~\eqref{eq:proj_update} only depends on neighboring sites \(i\) and \(i+1\), one can postpone the update of a matrix-product state until after updating all projectors between sites \(i\) and \(i+1\).
This postponement reduces the complexity scaling on the depth \(\tau \) from quadratic to linear.
Simultaneously, the acceptance rate decreases, and thus the autocorrelation of samples increases.
For the reduction of the scaling, we need a solution to the increase in the autocorrelation.
The treatment of random modes also has room for improvement.
The Kraus-Cirac decomposition increases the total depth of a unitary circuit.
In addition, the decomposition cannot be used for systems whose local Hilbert space dimension is larger than two.
The smarter treatment of random modes would considerably improve the efficiency of the TNMC approach.

As we observed in the Hamiltonian and the Floquet dynamics cases, adequate projectors depend on the characters of unitary evolution.
To increase the performance of the TNMC approach, one should choose problem-specific projectors, because ``there is no free lunch''.
Even for the TEBD-based projectors, there are some parameters, such as the threshold for computing the inverse or singular vectors for random modes.
One can use any weight function.
Since any projectors work in principle as long as the projectors satisfy the identity condition~\eqref{eq:identity}, the proposed TNMC approach intrinsically possesses arbitrariness.
We only presented working examples from many options of projectors.
A guiding principle for constructing projectors, other than testing them out, would be required.

As shown in this paper, the proposed TNMC approach can reproduce unbiased results within statistical errors, even with a bond dimension insufficient for the TEBD algorithms.
Consequently, we succeeded in implementing a classical algorithm for unitary evolution with numerically evaluable statistical errors, not truncation errors.
Such a tensor-network approach with statistical errors can be a new direction for improving the classical simulatability of unitary evolution.

\begin{acknowledgments}
    We thank X. Zhao for the fruitful discussions.
    This work was financially supported, by JSPS KAKENHI Grant No.~24K00543, by the Center of Innovation for Sustainable Quantum AI (SQAI), JST Grant No.~JPMJPF2221, and by JST CREST No.~JPMJCR24I1.
\end{acknowledgments}
\section*{Data availability}
The data that support the findings of this paper are openly available from https://doi.org/10.5281/zenodo.21465686.

The tensor-network simulations in this study were performed with ITensor library~\cite{fishman_itensor_2022}.
The code for the tensor-network Monte Carlo approach used in this paper is openly available from https://github.com/ShimpeiGoto/TEBD-TNMC.
\bibliographystyle{apsrev4-2} 

\begin{thebibliography}{55}%
\makeatletter
\providecommand \@ifxundefined [1]{%
 \@ifx{#1\undefined}
}%
\providecommand \@ifnum [1]{%
 \ifnum #1\expandafter \@firstoftwo
 \else \expandafter \@secondoftwo
 \fi
}%
\providecommand \@ifx [1]{%
 \ifx #1\expandafter \@firstoftwo
 \else \expandafter \@secondoftwo
 \fi
}%
\providecommand \natexlab [1]{#1}%
\providecommand \enquote  [1]{``#1''}%
\providecommand \bibnamefont  [1]{#1}%
\providecommand \bibfnamefont [1]{#1}%
\providecommand \citenamefont [1]{#1}%
\providecommand \href@noop [0]{\@secondoftwo}%
\providecommand \href [0]{\begingroup \@sanitize@url \@href}%
\providecommand \@href[1]{\@@startlink{#1}\@@href}%
\providecommand \@@href[1]{\endgroup#1\@@endlink}%
\providecommand \@sanitize@url [0]{\catcode `\\12\catcode `\$12\catcode `\&12\catcode `\#12\catcode `\^12\catcode `\_12\catcode `\%12\relax}%
\providecommand \@@startlink[1]{}%
\providecommand \@@endlink[0]{}%
\providecommand \url  [0]{\begingroup\@sanitize@url \@url }%
\providecommand \@url [1]{\endgroup\@href {#1}{\urlprefix }}%
\providecommand \urlprefix  [0]{URL }%
\providecommand \Eprint [0]{\href }%
\providecommand \doibase [0]{https://doi.org/}%
\providecommand \selectlanguage [0]{\@gobble}%
\providecommand \bibinfo  [0]{\@secondoftwo}%
\providecommand \bibfield  [0]{\@secondoftwo}%
\providecommand \translation [1]{[#1]}%
\providecommand \BibitemOpen [0]{}%
\providecommand \bibitemStop [0]{}%
\providecommand \bibitemNoStop [0]{.\EOS\space}%
\providecommand \EOS [0]{\spacefactor3000\relax}%
\providecommand \BibitemShut  [1]{\csname bibitem#1\endcsname}%
\let\auto@bib@innerbib\@empty
\bibitem [{\citenamefont {White}\ and\ \citenamefont {Feiguin}(2004)}]{white_real-time_2004}%
  \BibitemOpen
  \bibfield  {author} {\bibinfo {author} {\bibfnamefont {S.~R.}\ \bibnamefont {White}}\ and\ \bibinfo {author} {\bibfnamefont {A.~E.}\ \bibnamefont {Feiguin}},\ }\href {https://doi.org/10.1103/PhysRevLett.93.076401} {\bibfield  {journal} {\bibinfo  {journal} {Phys. Rev. Lett.}\ }\textbf {\bibinfo {volume} {93}},\ \bibinfo {pages} {076401} (\bibinfo {year} {2004})}\BibitemShut {NoStop}%
\bibitem [{\citenamefont {Baez}\ \emph {et~al.}(2020)\citenamefont {Baez}, \citenamefont {Goihl}, \citenamefont {Haferkamp}, \citenamefont {{Bermejo-Vega}}, \citenamefont {Gluza},\ and\ \citenamefont {Eisert}}]{baez_dynamical_2020}%
  \BibitemOpen
  \bibfield  {author} {\bibinfo {author} {\bibfnamefont {M.~L.}\ \bibnamefont {Baez}}, \bibinfo {author} {\bibfnamefont {M.}~\bibnamefont {Goihl}}, \bibinfo {author} {\bibfnamefont {J.}~\bibnamefont {Haferkamp}}, \bibinfo {author} {\bibfnamefont {J.}~\bibnamefont {{Bermejo-Vega}}}, \bibinfo {author} {\bibfnamefont {M.}~\bibnamefont {Gluza}},\ and\ \bibinfo {author} {\bibfnamefont {J.}~\bibnamefont {Eisert}},\ }\href {https://doi.org/10.1073/pnas.2006103117} {\bibfield  {journal} {\bibinfo  {journal} {Proceedings of the National Academy of Sciences}\ }\textbf {\bibinfo {volume} {117}},\ \bibinfo {pages} {26123} (\bibinfo {year} {2020})}\BibitemShut {NoStop}%
\bibitem [{\citenamefont {Gartner}\ \emph {et~al.}(2022)\citenamefont {Gartner}, \citenamefont {Mazzanti},\ and\ \citenamefont {Zillich}}]{gartner_time-dependent_2022}%
  \BibitemOpen
  \bibfield  {author} {\bibinfo {author} {\bibfnamefont {M.}~\bibnamefont {Gartner}}, \bibinfo {author} {\bibfnamefont {F.}~\bibnamefont {Mazzanti}},\ and\ \bibinfo {author} {\bibfnamefont {R.}~\bibnamefont {Zillich}},\ }\href {https://doi.org/10.21468/SciPostPhys.13.2.025} {\bibfield  {journal} {\bibinfo  {journal} {SciPost Physics}\ }\textbf {\bibinfo {volume} {13}},\ \bibinfo {pages} {025} (\bibinfo {year} {2022})}\BibitemShut {NoStop}%
\bibitem [{\citenamefont {Blais}\ \emph {et~al.}(2021)\citenamefont {Blais}, \citenamefont {Grimsmo}, \citenamefont {Girvin},\ and\ \citenamefont {Wallraff}}]{blais_circuit_2021}%
  \BibitemOpen
  \bibfield  {author} {\bibinfo {author} {\bibfnamefont {A.}~\bibnamefont {Blais}}, \bibinfo {author} {\bibfnamefont {A.~L.}\ \bibnamefont {Grimsmo}}, \bibinfo {author} {\bibfnamefont {S.~M.}\ \bibnamefont {Girvin}},\ and\ \bibinfo {author} {\bibfnamefont {A.}~\bibnamefont {Wallraff}},\ }\href {https://doi.org/10.1103/RevModPhys.93.025005} {\bibfield  {journal} {\bibinfo  {journal} {Rev. Mod. Phys.}\ }\textbf {\bibinfo {volume} {93}},\ \bibinfo {pages} {025005} (\bibinfo {year} {2021})}\BibitemShut {NoStop}%
\bibitem [{\citenamefont {Koch}\ \emph {et~al.}(2007)\citenamefont {Koch}, \citenamefont {Yu}, \citenamefont {Gambetta}, \citenamefont {Houck}, \citenamefont {Schuster}, \citenamefont {Majer}, \citenamefont {Blais}, \citenamefont {Devoret}, \citenamefont {Girvin},\ and\ \citenamefont {Schoelkopf}}]{koch_charge-insensitive_2007}%
  \BibitemOpen
  \bibfield  {author} {\bibinfo {author} {\bibfnamefont {J.}~\bibnamefont {Koch}}, \bibinfo {author} {\bibfnamefont {T.~M.}\ \bibnamefont {Yu}}, \bibinfo {author} {\bibfnamefont {J.}~\bibnamefont {Gambetta}}, \bibinfo {author} {\bibfnamefont {A.~A.}\ \bibnamefont {Houck}}, \bibinfo {author} {\bibfnamefont {D.~I.}\ \bibnamefont {Schuster}}, \bibinfo {author} {\bibfnamefont {J.}~\bibnamefont {Majer}}, \bibinfo {author} {\bibfnamefont {A.}~\bibnamefont {Blais}}, \bibinfo {author} {\bibfnamefont {M.~H.}\ \bibnamefont {Devoret}}, \bibinfo {author} {\bibfnamefont {S.~M.}\ \bibnamefont {Girvin}},\ and\ \bibinfo {author} {\bibfnamefont {R.~J.}\ \bibnamefont {Schoelkopf}},\ }\href {https://doi.org/10.1103/PhysRevA.76.042319} {\bibfield  {journal} {\bibinfo  {journal} {Phys. Rev. A}\ }\textbf {\bibinfo {volume} {76}},\ \bibinfo {pages} {042319} (\bibinfo {year} {2007})}\BibitemShut {NoStop}%
\bibitem [{\citenamefont {Schreier}\ \emph {et~al.}(2008)\citenamefont {Schreier}, \citenamefont {Houck}, \citenamefont {Koch}, \citenamefont {Schuster}, \citenamefont {Johnson}, \citenamefont {Chow}, \citenamefont {Gambetta}, \citenamefont {Majer}, \citenamefont {Frunzio}, \citenamefont {Devoret}, \citenamefont {Girvin},\ and\ \citenamefont {Schoelkopf}}]{schreier_suppressing_2008}%
  \BibitemOpen
  \bibfield  {author} {\bibinfo {author} {\bibfnamefont {J.~A.}\ \bibnamefont {Schreier}}, \bibinfo {author} {\bibfnamefont {A.~A.}\ \bibnamefont {Houck}}, \bibinfo {author} {\bibfnamefont {J.}~\bibnamefont {Koch}}, \bibinfo {author} {\bibfnamefont {D.~I.}\ \bibnamefont {Schuster}}, \bibinfo {author} {\bibfnamefont {B.~R.}\ \bibnamefont {Johnson}}, \bibinfo {author} {\bibfnamefont {J.~M.}\ \bibnamefont {Chow}}, \bibinfo {author} {\bibfnamefont {J.~M.}\ \bibnamefont {Gambetta}}, \bibinfo {author} {\bibfnamefont {J.}~\bibnamefont {Majer}}, \bibinfo {author} {\bibfnamefont {L.}~\bibnamefont {Frunzio}}, \bibinfo {author} {\bibfnamefont {M.~H.}\ \bibnamefont {Devoret}}, \bibinfo {author} {\bibfnamefont {S.~M.}\ \bibnamefont {Girvin}},\ and\ \bibinfo {author} {\bibfnamefont {R.~J.}\ \bibnamefont {Schoelkopf}},\ }\href {https://doi.org/10.1103/PhysRevB.77.180502} {\bibfield  {journal} {\bibinfo  {journal} {Phys. Rev. B}\ }\textbf {\bibinfo {volume} {77}},\ \bibinfo {pages} {180502} (\bibinfo {year} {2008})}\BibitemShut {NoStop}%
\bibitem [{\citenamefont {Kim}\ \emph {et~al.}(2023)\citenamefont {Kim}, \citenamefont {Eddins}, \citenamefont {Anand}, \citenamefont {Wei}, \citenamefont {{van den Berg}}, \citenamefont {Rosenblatt}, \citenamefont {Nayfeh}, \citenamefont {Wu}, \citenamefont {Zaletel}, \citenamefont {Temme},\ and\ \citenamefont {Kandala}}]{kim_evidence_2023}%
  \BibitemOpen
  \bibfield  {author} {\bibinfo {author} {\bibfnamefont {Y.}~\bibnamefont {Kim}}, \bibinfo {author} {\bibfnamefont {A.}~\bibnamefont {Eddins}}, \bibinfo {author} {\bibfnamefont {S.}~\bibnamefont {Anand}}, \bibinfo {author} {\bibfnamefont {K.~X.}\ \bibnamefont {Wei}}, \bibinfo {author} {\bibfnamefont {E.}~\bibnamefont {{van den Berg}}}, \bibinfo {author} {\bibfnamefont {S.}~\bibnamefont {Rosenblatt}}, \bibinfo {author} {\bibfnamefont {H.}~\bibnamefont {Nayfeh}}, \bibinfo {author} {\bibfnamefont {Y.}~\bibnamefont {Wu}}, \bibinfo {author} {\bibfnamefont {M.}~\bibnamefont {Zaletel}}, \bibinfo {author} {\bibfnamefont {K.}~\bibnamefont {Temme}},\ and\ \bibinfo {author} {\bibfnamefont {A.}~\bibnamefont {Kandala}},\ }\href {https://doi.org/10.1038/s41586-023-06096-3} {\bibfield  {journal} {\bibinfo  {journal} {Nature}\ }\textbf {\bibinfo {volume} {618}},\ \bibinfo {pages} {500} (\bibinfo {year} {2023})}\BibitemShut {NoStop}%
\bibitem [{\citenamefont {Bruzewicz}\ \emph {et~al.}(2019)\citenamefont {Bruzewicz}, \citenamefont {Chiaverini}, \citenamefont {McConnell},\ and\ \citenamefont {Sage}}]{bruzewicz_trapped-ion_2019}%
  \BibitemOpen
  \bibfield  {author} {\bibinfo {author} {\bibfnamefont {C.~D.}\ \bibnamefont {Bruzewicz}}, \bibinfo {author} {\bibfnamefont {J.}~\bibnamefont {Chiaverini}}, \bibinfo {author} {\bibfnamefont {R.}~\bibnamefont {McConnell}},\ and\ \bibinfo {author} {\bibfnamefont {J.~M.}\ \bibnamefont {Sage}},\ }\href {https://doi.org/10.1063/1.5088164} {\bibfield  {journal} {\bibinfo  {journal} {Appl. Phys. Rev.}\ }\textbf {\bibinfo {volume} {6}},\ \bibinfo {pages} {021314} (\bibinfo {year} {2019})}\BibitemShut {NoStop}%
\bibitem [{\citenamefont {Cirac}\ and\ \citenamefont {Zoller}(1995)}]{cirac_quantum_1995}%
  \BibitemOpen
  \bibfield  {author} {\bibinfo {author} {\bibfnamefont {J.~I.}\ \bibnamefont {Cirac}}\ and\ \bibinfo {author} {\bibfnamefont {P.}~\bibnamefont {Zoller}},\ }\href {https://doi.org/10.1103/PhysRevLett.74.4091} {\bibfield  {journal} {\bibinfo  {journal} {Phys. Rev. Lett.}\ }\textbf {\bibinfo {volume} {74}},\ \bibinfo {pages} {4091} (\bibinfo {year} {1995})}\BibitemShut {NoStop}%
\bibitem [{\citenamefont {Sackett}\ \emph {et~al.}(2000)\citenamefont {Sackett}, \citenamefont {Kielpinski}, \citenamefont {King}, \citenamefont {Langer}, \citenamefont {Meyer}, \citenamefont {Myatt}, \citenamefont {Rowe}, \citenamefont {Turchette}, \citenamefont {Itano}, \citenamefont {Wineland},\ and\ \citenamefont {Monroe}}]{sackett_experimental_2000}%
  \BibitemOpen
  \bibfield  {author} {\bibinfo {author} {\bibfnamefont {C.~A.}\ \bibnamefont {Sackett}}, \bibinfo {author} {\bibfnamefont {D.}~\bibnamefont {Kielpinski}}, \bibinfo {author} {\bibfnamefont {B.~E.}\ \bibnamefont {King}}, \bibinfo {author} {\bibfnamefont {C.}~\bibnamefont {Langer}}, \bibinfo {author} {\bibfnamefont {V.}~\bibnamefont {Meyer}}, \bibinfo {author} {\bibfnamefont {C.~J.}\ \bibnamefont {Myatt}}, \bibinfo {author} {\bibfnamefont {M.}~\bibnamefont {Rowe}}, \bibinfo {author} {\bibfnamefont {Q.~A.}\ \bibnamefont {Turchette}}, \bibinfo {author} {\bibfnamefont {W.~M.}\ \bibnamefont {Itano}}, \bibinfo {author} {\bibfnamefont {D.~J.}\ \bibnamefont {Wineland}},\ and\ \bibinfo {author} {\bibfnamefont {C.}~\bibnamefont {Monroe}},\ }\href {https://doi.org/10.1038/35005011} {\bibfield  {journal} {\bibinfo  {journal} {Nature}\ }\textbf {\bibinfo {volume} {404}},\ \bibinfo {pages} {256} (\bibinfo {year} {2000})}\BibitemShut {NoStop}%
\bibitem [{\citenamefont {Franke}\ \emph {et~al.}(2023)\citenamefont {Franke}, \citenamefont {Muleady}, \citenamefont {Kaubruegger}, \citenamefont {Kranzl}, \citenamefont {Blatt}, \citenamefont {Rey}, \citenamefont {Joshi},\ and\ \citenamefont {Roos}}]{franke_quantum-enhanced_2023}%
  \BibitemOpen
  \bibfield  {author} {\bibinfo {author} {\bibfnamefont {J.}~\bibnamefont {Franke}}, \bibinfo {author} {\bibfnamefont {S.~R.}\ \bibnamefont {Muleady}}, \bibinfo {author} {\bibfnamefont {R.}~\bibnamefont {Kaubruegger}}, \bibinfo {author} {\bibfnamefont {F.}~\bibnamefont {Kranzl}}, \bibinfo {author} {\bibfnamefont {R.}~\bibnamefont {Blatt}}, \bibinfo {author} {\bibfnamefont {A.~M.}\ \bibnamefont {Rey}}, \bibinfo {author} {\bibfnamefont {M.~K.}\ \bibnamefont {Joshi}},\ and\ \bibinfo {author} {\bibfnamefont {C.~F.}\ \bibnamefont {Roos}},\ }\href {https://doi.org/10.1038/s41586-023-06472-z} {\bibfield  {journal} {\bibinfo  {journal} {Nature}\ }\textbf {\bibinfo {volume} {621}},\ \bibinfo {pages} {740} (\bibinfo {year} {2023})}\BibitemShut {NoStop}%
\bibitem [{\citenamefont {Saffman}\ \emph {et~al.}(2010)\citenamefont {Saffman}, \citenamefont {Walker},\ and\ \citenamefont {M{\o}lmer}}]{saffman_quantum_2010}%
  \BibitemOpen
  \bibfield  {author} {\bibinfo {author} {\bibfnamefont {M.}~\bibnamefont {Saffman}}, \bibinfo {author} {\bibfnamefont {T.~G.}\ \bibnamefont {Walker}},\ and\ \bibinfo {author} {\bibfnamefont {K.}~\bibnamefont {M{\o}lmer}},\ }\href {https://doi.org/10.1103/RevModPhys.82.2313} {\bibfield  {journal} {\bibinfo  {journal} {Rev. Mod. Phys.}\ }\textbf {\bibinfo {volume} {82}},\ \bibinfo {pages} {2313} (\bibinfo {year} {2010})}\BibitemShut {NoStop}%
\bibitem [{\citenamefont {Maller}\ \emph {et~al.}(2015)\citenamefont {Maller}, \citenamefont {Lichtman}, \citenamefont {Xia}, \citenamefont {Sun}, \citenamefont {Piotrowicz}, \citenamefont {Carr}, \citenamefont {Isenhower},\ and\ \citenamefont {Saffman}}]{maller_rydberg-blockade_2015}%
  \BibitemOpen
  \bibfield  {author} {\bibinfo {author} {\bibfnamefont {K.~M.}\ \bibnamefont {Maller}}, \bibinfo {author} {\bibfnamefont {M.~T.}\ \bibnamefont {Lichtman}}, \bibinfo {author} {\bibfnamefont {T.}~\bibnamefont {Xia}}, \bibinfo {author} {\bibfnamefont {Y.}~\bibnamefont {Sun}}, \bibinfo {author} {\bibfnamefont {M.~J.}\ \bibnamefont {Piotrowicz}}, \bibinfo {author} {\bibfnamefont {A.~W.}\ \bibnamefont {Carr}}, \bibinfo {author} {\bibfnamefont {L.}~\bibnamefont {Isenhower}},\ and\ \bibinfo {author} {\bibfnamefont {M.}~\bibnamefont {Saffman}},\ }\href {https://doi.org/10.1103/PhysRevA.92.022336} {\bibfield  {journal} {\bibinfo  {journal} {Phys. Rev. A}\ }\textbf {\bibinfo {volume} {92}},\ \bibinfo {pages} {022336} (\bibinfo {year} {2015})}\BibitemShut {NoStop}%
\bibitem [{\citenamefont {Manetsch}\ \emph {et~al.}(2025)\citenamefont {Manetsch}, \citenamefont {Nomura}, \citenamefont {Bataille}, \citenamefont {Lv}, \citenamefont {Leung},\ and\ \citenamefont {Endres}}]{manetsch_tweezer_2025}%
  \BibitemOpen
  \bibfield  {author} {\bibinfo {author} {\bibfnamefont {H.~J.}\ \bibnamefont {Manetsch}}, \bibinfo {author} {\bibfnamefont {G.}~\bibnamefont {Nomura}}, \bibinfo {author} {\bibfnamefont {E.}~\bibnamefont {Bataille}}, \bibinfo {author} {\bibfnamefont {X.}~\bibnamefont {Lv}}, \bibinfo {author} {\bibfnamefont {K.~H.}\ \bibnamefont {Leung}},\ and\ \bibinfo {author} {\bibfnamefont {M.}~\bibnamefont {Endres}},\ }\href {https://doi.org/10.1038/s41586-025-09641-4} {\bibfield  {journal} {\bibinfo  {journal} {Nature}\ }\textbf {\bibinfo {volume} {647}},\ \bibinfo {pages} {60} (\bibinfo {year} {2025})}\BibitemShut {NoStop}%
\bibitem [{\citenamefont {Greiner}\ \emph {et~al.}(2002)\citenamefont {Greiner}, \citenamefont {Mandel}, \citenamefont {H{\"a}nsch},\ and\ \citenamefont {Bloch}}]{greiner_collapse_2002}%
  \BibitemOpen
  \bibfield  {author} {\bibinfo {author} {\bibfnamefont {M.}~\bibnamefont {Greiner}}, \bibinfo {author} {\bibfnamefont {O.}~\bibnamefont {Mandel}}, \bibinfo {author} {\bibfnamefont {T.~W.}\ \bibnamefont {H{\"a}nsch}},\ and\ \bibinfo {author} {\bibfnamefont {I.}~\bibnamefont {Bloch}},\ }\href {https://doi.org/10.1038/nature00968} {\bibfield  {journal} {\bibinfo  {journal} {Nature}\ }\textbf {\bibinfo {volume} {419}},\ \bibinfo {pages} {51} (\bibinfo {year} {2002})}\BibitemShut {NoStop}%
\bibitem [{\citenamefont {Trotzky}\ \emph {et~al.}(2012)\citenamefont {Trotzky}, \citenamefont {Chen}, \citenamefont {Flesch}, \citenamefont {McCulloch}, \citenamefont {Schollw{\"o}ck}, \citenamefont {Eisert},\ and\ \citenamefont {Bloch}}]{trotzky_probing_2012}%
  \BibitemOpen
  \bibfield  {author} {\bibinfo {author} {\bibfnamefont {S.}~\bibnamefont {Trotzky}}, \bibinfo {author} {\bibfnamefont {Y.-A.}\ \bibnamefont {Chen}}, \bibinfo {author} {\bibfnamefont {A.}~\bibnamefont {Flesch}}, \bibinfo {author} {\bibfnamefont {I.~P.}\ \bibnamefont {McCulloch}}, \bibinfo {author} {\bibfnamefont {U.}~\bibnamefont {Schollw{\"o}ck}}, \bibinfo {author} {\bibfnamefont {J.}~\bibnamefont {Eisert}},\ and\ \bibinfo {author} {\bibfnamefont {I.}~\bibnamefont {Bloch}},\ }\href {https://doi.org/10.1038/nphys2232} {\bibfield  {journal} {\bibinfo  {journal} {Nat Phys}\ }\textbf {\bibinfo {volume} {8}},\ \bibinfo {pages} {325} (\bibinfo {year} {2012})}\BibitemShut {NoStop}%
\bibitem [{\citenamefont {Bernien}\ \emph {et~al.}(2017)\citenamefont {Bernien}, \citenamefont {Schwartz}, \citenamefont {Keesling}, \citenamefont {Levine}, \citenamefont {Omran}, \citenamefont {Pichler}, \citenamefont {Choi}, \citenamefont {Zibrov}, \citenamefont {Endres}, \citenamefont {Greiner}, \citenamefont {Vuleti{\'c}},\ and\ \citenamefont {Lukin}}]{bernien_probing_2017}%
  \BibitemOpen
  \bibfield  {author} {\bibinfo {author} {\bibfnamefont {H.}~\bibnamefont {Bernien}}, \bibinfo {author} {\bibfnamefont {S.}~\bibnamefont {Schwartz}}, \bibinfo {author} {\bibfnamefont {A.}~\bibnamefont {Keesling}}, \bibinfo {author} {\bibfnamefont {H.}~\bibnamefont {Levine}}, \bibinfo {author} {\bibfnamefont {A.}~\bibnamefont {Omran}}, \bibinfo {author} {\bibfnamefont {H.}~\bibnamefont {Pichler}}, \bibinfo {author} {\bibfnamefont {S.}~\bibnamefont {Choi}}, \bibinfo {author} {\bibfnamefont {A.~S.}\ \bibnamefont {Zibrov}}, \bibinfo {author} {\bibfnamefont {M.}~\bibnamefont {Endres}}, \bibinfo {author} {\bibfnamefont {M.}~\bibnamefont {Greiner}}, \bibinfo {author} {\bibfnamefont {V.}~\bibnamefont {Vuleti{\'c}}},\ and\ \bibinfo {author} {\bibfnamefont {M.~D.}\ \bibnamefont {Lukin}},\ }\href {https://doi.org/10.1038/nature24622} {\bibfield  {journal} {\bibinfo  {journal} {Nature}\ }\textbf {\bibinfo {volume} {551}},\ \bibinfo {pages} {579} (\bibinfo {year} {2017})}\BibitemShut {NoStop}%
\bibitem [{\citenamefont {Takasu}\ \emph {et~al.}(2020)\citenamefont {Takasu}, \citenamefont {Yagami}, \citenamefont {Asaka}, \citenamefont {Fukushima}, \citenamefont {Nagao}, \citenamefont {Goto}, \citenamefont {Danshita},\ and\ \citenamefont {Takahashi}}]{takasu_energy_2020-1}%
  \BibitemOpen
  \bibfield  {author} {\bibinfo {author} {\bibfnamefont {Y.}~\bibnamefont {Takasu}}, \bibinfo {author} {\bibfnamefont {T.}~\bibnamefont {Yagami}}, \bibinfo {author} {\bibfnamefont {H.}~\bibnamefont {Asaka}}, \bibinfo {author} {\bibfnamefont {Y.}~\bibnamefont {Fukushima}}, \bibinfo {author} {\bibfnamefont {K.}~\bibnamefont {Nagao}}, \bibinfo {author} {\bibfnamefont {S.}~\bibnamefont {Goto}}, \bibinfo {author} {\bibfnamefont {I.}~\bibnamefont {Danshita}},\ and\ \bibinfo {author} {\bibfnamefont {Y.}~\bibnamefont {Takahashi}},\ }\href {https://doi.org/10.1126/sciadv.aba9255} {\bibfield  {journal} {\bibinfo  {journal} {Science Advances}\ }\textbf {\bibinfo {volume} {6}},\ \bibinfo {pages} {eaba9255} (\bibinfo {year} {2020})}\BibitemShut {NoStop}%
\bibitem [{\citenamefont {Honda}\ \emph {et~al.}(2025)\citenamefont {Honda}, \citenamefont {Takasu}, \citenamefont {Goto}, \citenamefont {Kazuta}, \citenamefont {Kunimi}, \citenamefont {Danshita},\ and\ \citenamefont {Takahashi}}]{honda_observation_2025}%
  \BibitemOpen
  \bibfield  {author} {\bibinfo {author} {\bibfnamefont {K.}~\bibnamefont {Honda}}, \bibinfo {author} {\bibfnamefont {Y.}~\bibnamefont {Takasu}}, \bibinfo {author} {\bibfnamefont {S.}~\bibnamefont {Goto}}, \bibinfo {author} {\bibfnamefont {H.}~\bibnamefont {Kazuta}}, \bibinfo {author} {\bibfnamefont {M.}~\bibnamefont {Kunimi}}, \bibinfo {author} {\bibfnamefont {I.}~\bibnamefont {Danshita}},\ and\ \bibinfo {author} {\bibfnamefont {Y.}~\bibnamefont {Takahashi}},\ }\href {https://doi.org/10.1126/sciadv.adv3255} {\bibfield  {journal} {\bibinfo  {journal} {Science Advances}\ }\textbf {\bibinfo {volume} {11}},\ \bibinfo {pages} {eadv3255} (\bibinfo {year} {2025})}\BibitemShut {NoStop}%
\bibitem [{\citenamefont {Schollw{\"o}ck}(2011)}]{schollwock_density-matrix_2011}%
  \BibitemOpen
  \bibfield  {author} {\bibinfo {author} {\bibfnamefont {U.}~\bibnamefont {Schollw{\"o}ck}},\ }\href {https://doi.org/10.1016/j.aop.2010.09.012} {\bibfield  {journal} {\bibinfo  {journal} {Annals of Physics}\ }\textbf {\bibinfo {volume} {326}},\ \bibinfo {pages} {96} (\bibinfo {year} {2011})}\BibitemShut {NoStop}%
\bibitem [{\citenamefont {Cirac}\ \emph {et~al.}(2021)\citenamefont {Cirac}, \citenamefont {{P{\'e}rez-Garc{\'i}a}}, \citenamefont {Schuch},\ and\ \citenamefont {Verstraete}}]{cirac_matrix_2021}%
  \BibitemOpen
  \bibfield  {author} {\bibinfo {author} {\bibfnamefont {J.~I.}\ \bibnamefont {Cirac}}, \bibinfo {author} {\bibfnamefont {D.}~\bibnamefont {{P{\'e}rez-Garc{\'i}a}}}, \bibinfo {author} {\bibfnamefont {N.}~\bibnamefont {Schuch}},\ and\ \bibinfo {author} {\bibfnamefont {F.}~\bibnamefont {Verstraete}},\ }\href {https://doi.org/10.1103/RevModPhys.93.045003} {\bibfield  {journal} {\bibinfo  {journal} {Rev. Mod. Phys.}\ }\textbf {\bibinfo {volume} {93}},\ \bibinfo {pages} {045003} (\bibinfo {year} {2021})}\BibitemShut {NoStop}%
\bibitem [{\citenamefont {Vidal}(2003)}]{vidal_efficient_2003}%
  \BibitemOpen
  \bibfield  {author} {\bibinfo {author} {\bibfnamefont {G.}~\bibnamefont {Vidal}},\ }\href {https://doi.org/10.1103/PhysRevLett.91.147902} {\bibfield  {journal} {\bibinfo  {journal} {Phys. Rev. Lett.}\ }\textbf {\bibinfo {volume} {91}},\ \bibinfo {pages} {147902} (\bibinfo {year} {2003})}\BibitemShut {NoStop}%
\bibitem [{\citenamefont {Vidal}(2004)}]{vidal_efficient_2004}%
  \BibitemOpen
  \bibfield  {author} {\bibinfo {author} {\bibfnamefont {G.}~\bibnamefont {Vidal}},\ }\href {https://doi.org/10.1103/PhysRevLett.93.040502} {\bibfield  {journal} {\bibinfo  {journal} {Phys. Rev. Lett.}\ }\textbf {\bibinfo {volume} {93}},\ \bibinfo {pages} {040502} (\bibinfo {year} {2004})}\BibitemShut {NoStop}%
\bibitem [{\citenamefont {Daley}\ \emph {et~al.}(2004)\citenamefont {Daley}, \citenamefont {Kollath}, \citenamefont {Schollw{\"o}ck},\ and\ \citenamefont {Vidal}}]{daley_time-dependent_2004}%
  \BibitemOpen
  \bibfield  {author} {\bibinfo {author} {\bibfnamefont {A.~J.}\ \bibnamefont {Daley}}, \bibinfo {author} {\bibfnamefont {C.}~\bibnamefont {Kollath}}, \bibinfo {author} {\bibfnamefont {U.}~\bibnamefont {Schollw{\"o}ck}},\ and\ \bibinfo {author} {\bibfnamefont {G.}~\bibnamefont {Vidal}},\ }\href {https://doi.org/10.1088/1742-5468/2004/04/P04005} {\bibfield  {journal} {\bibinfo  {journal} {J. Stat. Mech.}\ }\textbf {\bibinfo {volume} {2004}},\ \bibinfo {pages} {P04005} (\bibinfo {year} {2004})}\BibitemShut {NoStop}%
\bibitem [{\citenamefont {{Garc{\'i}a-Ripoll}}(2006)}]{garcia-ripoll_time_2006}%
  \BibitemOpen
  \bibfield  {author} {\bibinfo {author} {\bibfnamefont {J.~J.}\ \bibnamefont {{Garc{\'i}a-Ripoll}}},\ }\href {https://doi.org/10.1088/1367-2630/8/12/305} {\bibfield  {journal} {\bibinfo  {journal} {New J. Phys.}\ }\textbf {\bibinfo {volume} {8}},\ \bibinfo {pages} {305} (\bibinfo {year} {2006})}\BibitemShut {NoStop}%
\bibitem [{\citenamefont {Wall}\ and\ \citenamefont {Carr}(2012)}]{wall_out--equilibrium_2012}%
  \BibitemOpen
  \bibfield  {author} {\bibinfo {author} {\bibfnamefont {M.~L.}\ \bibnamefont {Wall}}\ and\ \bibinfo {author} {\bibfnamefont {L.~D.}\ \bibnamefont {Carr}},\ }\href {https://doi.org/10.1088/1367-2630/14/12/125015} {\bibfield  {journal} {\bibinfo  {journal} {New J. Phys.}\ }\textbf {\bibinfo {volume} {14}},\ \bibinfo {pages} {125015} (\bibinfo {year} {2012})}\BibitemShut {NoStop}%
\bibitem [{\citenamefont {Haegeman}\ \emph {et~al.}(2016)\citenamefont {Haegeman}, \citenamefont {Lubich}, \citenamefont {Oseledets}, \citenamefont {Vandereycken},\ and\ \citenamefont {Verstraete}}]{haegeman_unifying_2016}%
  \BibitemOpen
  \bibfield  {author} {\bibinfo {author} {\bibfnamefont {J.}~\bibnamefont {Haegeman}}, \bibinfo {author} {\bibfnamefont {C.}~\bibnamefont {Lubich}}, \bibinfo {author} {\bibfnamefont {I.}~\bibnamefont {Oseledets}}, \bibinfo {author} {\bibfnamefont {B.}~\bibnamefont {Vandereycken}},\ and\ \bibinfo {author} {\bibfnamefont {F.}~\bibnamefont {Verstraete}},\ }\href {https://doi.org/10.1103/PhysRevB.94.165116} {\bibfield  {journal} {\bibinfo  {journal} {Phys. Rev. B}\ }\textbf {\bibinfo {volume} {94}},\ \bibinfo {pages} {165116} (\bibinfo {year} {2016})}\BibitemShut {NoStop}%
\bibitem [{\citenamefont {Li}\ \emph {et~al.}(2024)\citenamefont {Li}, \citenamefont {Gleis},\ and\ \citenamefont {{von Delft}}}]{li_time-dependent_2024}%
  \BibitemOpen
  \bibfield  {author} {\bibinfo {author} {\bibfnamefont {J.-W.}\ \bibnamefont {Li}}, \bibinfo {author} {\bibfnamefont {A.}~\bibnamefont {Gleis}},\ and\ \bibinfo {author} {\bibfnamefont {J.}~\bibnamefont {{von Delft}}},\ }\href {https://doi.org/10.1103/PhysRevLett.133.026401} {\bibfield  {journal} {\bibinfo  {journal} {Phys. Rev. Lett.}\ }\textbf {\bibinfo {volume} {133}},\ \bibinfo {pages} {026401} (\bibinfo {year} {2024})}\BibitemShut {NoStop}%
\bibitem [{\citenamefont {Zaletel}\ \emph {et~al.}(2015)\citenamefont {Zaletel}, \citenamefont {Mong}, \citenamefont {Karrasch}, \citenamefont {Moore},\ and\ \citenamefont {Pollmann}}]{zaletel_time-evolving_2015}%
  \BibitemOpen
  \bibfield  {author} {\bibinfo {author} {\bibfnamefont {M.~P.}\ \bibnamefont {Zaletel}}, \bibinfo {author} {\bibfnamefont {R.~S.~K.}\ \bibnamefont {Mong}}, \bibinfo {author} {\bibfnamefont {C.}~\bibnamefont {Karrasch}}, \bibinfo {author} {\bibfnamefont {J.~E.}\ \bibnamefont {Moore}},\ and\ \bibinfo {author} {\bibfnamefont {F.}~\bibnamefont {Pollmann}},\ }\href {https://doi.org/10.1103/PhysRevB.91.165112} {\bibfield  {journal} {\bibinfo  {journal} {Phys. Rev. B}\ }\textbf {\bibinfo {volume} {91}},\ \bibinfo {pages} {165112} (\bibinfo {year} {2015})}\BibitemShut {NoStop}%
\bibitem [{\citenamefont {Czarnik}\ and\ \citenamefont {Dziarmaga}(2018)}]{czarnik_time_2018}%
  \BibitemOpen
  \bibfield  {author} {\bibinfo {author} {\bibfnamefont {P.}~\bibnamefont {Czarnik}}\ and\ \bibinfo {author} {\bibfnamefont {J.}~\bibnamefont {Dziarmaga}},\ }\href {https://doi.org/10.1103/PhysRevB.98.045110} {\bibfield  {journal} {\bibinfo  {journal} {Phys. Rev. B}\ }\textbf {\bibinfo {volume} {98}},\ \bibinfo {pages} {045110} (\bibinfo {year} {2018})}\BibitemShut {NoStop}%
\bibitem [{\citenamefont {Czarnik}\ \emph {et~al.}(2019)\citenamefont {Czarnik}, \citenamefont {Dziarmaga},\ and\ \citenamefont {Corboz}}]{czarnik_time_2019}%
  \BibitemOpen
  \bibfield  {author} {\bibinfo {author} {\bibfnamefont {P.}~\bibnamefont {Czarnik}}, \bibinfo {author} {\bibfnamefont {J.}~\bibnamefont {Dziarmaga}},\ and\ \bibinfo {author} {\bibfnamefont {P.}~\bibnamefont {Corboz}},\ }\href {https://doi.org/10.1103/PhysRevB.99.035115} {\bibfield  {journal} {\bibinfo  {journal} {Phys. Rev. B}\ }\textbf {\bibinfo {volume} {99}},\ \bibinfo {pages} {035115} (\bibinfo {year} {2019})}\BibitemShut {NoStop}%
\bibitem [{\citenamefont {Ferris}(2015)}]{ferris_unbiased_2015}%
  \BibitemOpen
  \bibfield  {author} {\bibinfo {author} {\bibfnamefont {A.~J.}\ \bibnamefont {Ferris}},\ }\href {https://doi.org/10.48550/arXiv.1507.00767} {\bibinfo {title} {Unbiased {{Monte Carlo}} for the age of tensor networks}} (\bibinfo {year} {2015}),\ \Eprint {https://arxiv.org/abs/1507.00767} {arXiv:1507.00767 [cond-mat]} \BibitemShut {NoStop}%
\bibitem [{\citenamefont {Arai}\ \emph {et~al.}(2023)\citenamefont {Arai}, \citenamefont {Ohki}, \citenamefont {Takeda},\ and\ \citenamefont {Tomii}}]{arai_all-mode_2023}%
  \BibitemOpen
  \bibfield  {author} {\bibinfo {author} {\bibfnamefont {E.}~\bibnamefont {Arai}}, \bibinfo {author} {\bibfnamefont {H.}~\bibnamefont {Ohki}}, \bibinfo {author} {\bibfnamefont {S.}~\bibnamefont {Takeda}},\ and\ \bibinfo {author} {\bibfnamefont {M.}~\bibnamefont {Tomii}},\ }\href {https://doi.org/10.1103/PhysRevD.107.114515} {\bibfield  {journal} {\bibinfo  {journal} {Phys. Rev. D}\ }\textbf {\bibinfo {volume} {107}},\ \bibinfo {pages} {114515} (\bibinfo {year} {2023})}\BibitemShut {NoStop}%
\bibitem [{\citenamefont {Todo}(2024)}]{todo_markov_2024}%
  \BibitemOpen
  \bibfield  {author} {\bibinfo {author} {\bibfnamefont {S.}~\bibnamefont {Todo}},\ }\href {https://doi.org/10.48550/arXiv.2412.02974} {\bibinfo {title} {Markov {{Chain Monte Carlo}} in {{Tensor Network Representation}}}} (\bibinfo {year} {2024}),\ \Eprint {https://arxiv.org/abs/2412.02974} {arXiv:2412.02974 [cond-mat]} \BibitemShut {NoStop}%
\bibitem [{Note1()}]{Note1}%
  \BibitemOpen
  \bibinfo {note} {The term ``tensor network Monte Carlo'' is used in different contexts such as the variational Monte Carlo evaluation of tensor-network contraction or the proposal of spin configurations from tensor-network contraction.}\BibitemShut {Stop}%
\bibitem [{\citenamefont {Levin}\ and\ \citenamefont {Nave}(2007)}]{levin_tensor_2007}%
  \BibitemOpen
  \bibfield  {author} {\bibinfo {author} {\bibfnamefont {M.}~\bibnamefont {Levin}}\ and\ \bibinfo {author} {\bibfnamefont {C.~P.}\ \bibnamefont {Nave}},\ }\href {https://doi.org/10.1103/PhysRevLett.99.120601} {\bibfield  {journal} {\bibinfo  {journal} {Phys. Rev. Lett.}\ }\textbf {\bibinfo {volume} {99}},\ \bibinfo {pages} {120601} (\bibinfo {year} {2007})}\BibitemShut {NoStop}%
\bibitem [{\citenamefont {Morita}\ and\ \citenamefont {Kawashima}(2019)}]{morita_calculation_2019}%
  \BibitemOpen
  \bibfield  {author} {\bibinfo {author} {\bibfnamefont {S.}~\bibnamefont {Morita}}\ and\ \bibinfo {author} {\bibfnamefont {N.}~\bibnamefont {Kawashima}},\ }\href {https://doi.org/10.1016/j.cpc.2018.10.014} {\bibfield  {journal} {\bibinfo  {journal} {Computer Physics Communications}\ }\textbf {\bibinfo {volume} {236}},\ \bibinfo {pages} {65} (\bibinfo {year} {2019})}\BibitemShut {NoStop}%
\bibitem [{\citenamefont {Pan}\ and\ \citenamefont {Meng}(2024)}]{pan_sign_2024}%
  \BibitemOpen
  \bibfield  {author} {\bibinfo {author} {\bibfnamefont {G.}~\bibnamefont {Pan}}\ and\ \bibinfo {author} {\bibfnamefont {Z.~Y.}\ \bibnamefont {Meng}},\ }in\ \href {https://doi.org/10.1016/B978-0-323-90800-9.00095-0} {\emph {\bibinfo {booktitle} {Encyclopedia of {{Condensed Matter Physics}} ({{Second Edition}})}}},\ \bibinfo {editor} {edited by\ \bibinfo {editor} {\bibfnamefont {T.}~\bibnamefont {Chakraborty}}}\ (\bibinfo  {publisher} {Academic Press},\ \bibinfo {address} {Oxford},\ \bibinfo {year} {2024})\ pp.\ \bibinfo {pages} {879--893}\BibitemShut {NoStop}%
\bibitem [{\citenamefont {Bertini}\ \emph {et~al.}(2019{\natexlab{a}})\citenamefont {Bertini}, \citenamefont {Kos},\ and\ \citenamefont {Prosen}}]{bertini_exact_2019}%
  \BibitemOpen
  \bibfield  {author} {\bibinfo {author} {\bibfnamefont {B.}~\bibnamefont {Bertini}}, \bibinfo {author} {\bibfnamefont {P.}~\bibnamefont {Kos}},\ and\ \bibinfo {author} {\bibfnamefont {T.}~\bibnamefont {Prosen}},\ }\href {https://doi.org/10.1103/PhysRevLett.123.210601} {\bibfield  {journal} {\bibinfo  {journal} {Phys. Rev. Lett.}\ }\textbf {\bibinfo {volume} {123}},\ \bibinfo {pages} {210601} (\bibinfo {year} {2019}{\natexlab{a}})}\BibitemShut {NoStop}%
\bibitem [{\citenamefont {Bertini}\ \emph {et~al.}(2019{\natexlab{b}})\citenamefont {Bertini}, \citenamefont {Kos},\ and\ \citenamefont {Prosen}}]{bertini_entanglement_2019}%
  \BibitemOpen
  \bibfield  {author} {\bibinfo {author} {\bibfnamefont {B.}~\bibnamefont {Bertini}}, \bibinfo {author} {\bibfnamefont {P.}~\bibnamefont {Kos}},\ and\ \bibinfo {author} {\bibfnamefont {T.}~\bibnamefont {Prosen}},\ }\href {https://doi.org/10.1103/PhysRevX.9.021033} {\bibfield  {journal} {\bibinfo  {journal} {Phys. Rev. X}\ }\textbf {\bibinfo {volume} {9}},\ \bibinfo {pages} {021033} (\bibinfo {year} {2019}{\natexlab{b}})}\BibitemShut {NoStop}%
\bibitem [{\citenamefont {Piroli}\ \emph {et~al.}(2020)\citenamefont {Piroli}, \citenamefont {Bertini}, \citenamefont {Cirac},\ and\ \citenamefont {Prosen}}]{piroli_exact_2020}%
  \BibitemOpen
  \bibfield  {author} {\bibinfo {author} {\bibfnamefont {L.}~\bibnamefont {Piroli}}, \bibinfo {author} {\bibfnamefont {B.}~\bibnamefont {Bertini}}, \bibinfo {author} {\bibfnamefont {J.~I.}\ \bibnamefont {Cirac}},\ and\ \bibinfo {author} {\bibfnamefont {T.}~\bibnamefont {Prosen}},\ }\href {https://doi.org/10.1103/PhysRevB.101.094304} {\bibfield  {journal} {\bibinfo  {journal} {Phys. Rev. B}\ }\textbf {\bibinfo {volume} {101}},\ \bibinfo {pages} {094304} (\bibinfo {year} {2020})}\BibitemShut {NoStop}%
\bibitem [{Note2()}]{Note2}%
  \BibitemOpen
  \bibinfo {note} {Although we call the tensor \(\protect \bm {P}^{b, b^\prime }(\theta )\) a projector, this tensor does not need to be idempotent. The requirement for a ``projector'' in this paper is this identity condition.}\BibitemShut {Stop}%
\bibitem [{\citenamefont {Metropolis}\ \emph {et~al.}(1953)\citenamefont {Metropolis}, \citenamefont {Rosenbluth}, \citenamefont {Rosenbluth}, \citenamefont {Teller},\ and\ \citenamefont {Teller}}]{metropolis_equation_1953}%
  \BibitemOpen
  \bibfield  {author} {\bibinfo {author} {\bibfnamefont {N.}~\bibnamefont {Metropolis}}, \bibinfo {author} {\bibfnamefont {A.~W.}\ \bibnamefont {Rosenbluth}}, \bibinfo {author} {\bibfnamefont {M.~N.}\ \bibnamefont {Rosenbluth}}, \bibinfo {author} {\bibfnamefont {A.~H.}\ \bibnamefont {Teller}},\ and\ \bibinfo {author} {\bibfnamefont {E.}~\bibnamefont {Teller}},\ }\href {https://doi.org/10.1063/1.1699114} {\bibfield  {journal} {\bibinfo  {journal} {J. Chem. Phys.}\ }\textbf {\bibinfo {volume} {21}},\ \bibinfo {pages} {1087} (\bibinfo {year} {1953})}\BibitemShut {NoStop}%
\bibitem [{\citenamefont {Hastings}(1970)}]{hastings_monte_1970}%
  \BibitemOpen
  \bibfield  {author} {\bibinfo {author} {\bibfnamefont {W.~K.}\ \bibnamefont {Hastings}},\ }\href {https://doi.org/10.1093/biomet/57.1.97} {\bibfield  {journal} {\bibinfo  {journal} {Biometrika}\ }\textbf {\bibinfo {volume} {57}},\ \bibinfo {pages} {97} (\bibinfo {year} {1970})}\BibitemShut {NoStop}%
\bibitem [{Note3()}]{Note3}%
  \BibitemOpen
  \bibinfo {note} {If one computes the overlap \(\mathinner {\langle {\psi (\protect \bm {\theta })|\psi (\protect \bm {\theta })}\rangle }\) naively to evaluate the weight function, an additional factor \(N\) arises in the complexity. This additional factor can be eliminated by storing tensors appearing during the computation of the overlap as often performed in the computation of effective Hamiltonian in the density-matrix renormalization group approach~\cite {schollwock_density-matrix_2011}.}\BibitemShut {Stop}%
\bibitem [{\citenamefont {Mezzadri}(2007)}]{mezzadri_how_2007}%
  \BibitemOpen
  \bibfield  {author} {\bibinfo {author} {\bibfnamefont {F.}~\bibnamefont {Mezzadri}},\ }\href@noop {} {\bibfield  {journal} {\bibinfo  {journal} {NOTICES of the AMS}\ }\textbf {\bibinfo {volume} {54}},\ \bibinfo {pages} {592} (\bibinfo {year} {2007})}\BibitemShut {NoStop}%
\bibitem [{\citenamefont {Hastings}(2009)}]{hastings_light-cone_2009}%
  \BibitemOpen
  \bibfield  {author} {\bibinfo {author} {\bibfnamefont {M.~B.}\ \bibnamefont {Hastings}},\ }\href {https://doi.org/10.1063/1.3149556} {\bibfield  {journal} {\bibinfo  {journal} {Journal of Mathematical Physics}\ }\textbf {\bibinfo {volume} {50}},\ \bibinfo {pages} {095207} (\bibinfo {year} {2009})}\BibitemShut {NoStop}%
\bibitem [{\citenamefont {Golub}\ and\ \citenamefont {Loan}(2012)}]{golub_matrix_2012}%
  \BibitemOpen
  \bibfield  {author} {\bibinfo {author} {\bibfnamefont {G.~H.}\ \bibnamefont {Golub}}\ and\ \bibinfo {author} {\bibfnamefont {C.~F.~V.}\ \bibnamefont {Loan}},\ }\href@noop {} {\emph {\bibinfo {title} {Matrix {{Computations}}}}},\ \bibinfo {edition} {4th}\ ed.\ (\bibinfo  {publisher} {Johns Hopkins Univ Pr},\ \bibinfo {address} {Baltimore},\ \bibinfo {year} {2012})\BibitemShut {NoStop}%
\bibitem [{\citenamefont {Kraus}\ and\ \citenamefont {Cirac}(2001)}]{kraus_optimal_2001}%
  \BibitemOpen
  \bibfield  {author} {\bibinfo {author} {\bibfnamefont {B.}~\bibnamefont {Kraus}}\ and\ \bibinfo {author} {\bibfnamefont {J.~I.}\ \bibnamefont {Cirac}},\ }\href {https://doi.org/10.1103/PhysRevA.63.062309} {\bibfield  {journal} {\bibinfo  {journal} {Phys. Rev. A}\ }\textbf {\bibinfo {volume} {63}},\ \bibinfo {pages} {062309} (\bibinfo {year} {2001})}\BibitemShut {NoStop}%
\bibitem [{\citenamefont {Hatano}\ and\ \citenamefont {Suzuki}(2005)}]{hatano_finding_2005}%
  \BibitemOpen
  \bibfield  {author} {\bibinfo {author} {\bibfnamefont {N.}~\bibnamefont {Hatano}}\ and\ \bibinfo {author} {\bibfnamefont {M.}~\bibnamefont {Suzuki}},\ }in\ \href {https://doi.org/10.1007/11526216_2} {\emph {\bibinfo {booktitle} {Quantum {{Annealing}} and {{Other Optimization Methods}}}}},\ \bibinfo {series and number} {Lecture {{Notes}} in {{Physics}}},\ \bibinfo {editor} {edited by\ \bibinfo {editor} {\bibfnamefont {A.}~\bibnamefont {Das}}\ and\ \bibinfo {editor} {\bibfnamefont {B.}~\bibnamefont {K.~Chakrabarti}}}\ (\bibinfo  {publisher} {Springer},\ \bibinfo {address} {Berlin, Heidelberg},\ \bibinfo {year} {2005})\ pp.\ \bibinfo {pages} {37--68}\BibitemShut {NoStop}%
\bibitem [{\citenamefont {Sokal}(1997)}]{sokal_monte_1997}%
  \BibitemOpen
  \bibfield  {author} {\bibinfo {author} {\bibfnamefont {A.}~\bibnamefont {Sokal}},\ }in\ \href {https://doi.org/10.1007/978-1-4899-0319-8_6} {\emph {\bibinfo {booktitle} {Functional {{Integration}}: {{Basics}} and {{Applications}}}}},\ \bibinfo {series and number} {{{NATO ASI Series}}},\ \bibinfo {editor} {edited by\ \bibinfo {editor} {\bibfnamefont {C.}~\bibnamefont {{DeWitt-Morette}}}, \bibinfo {editor} {\bibfnamefont {P.}~\bibnamefont {Cartier}},\ and\ \bibinfo {editor} {\bibfnamefont {A.}~\bibnamefont {Folacci}}}\ (\bibinfo  {publisher} {Springer US},\ \bibinfo {address} {Boston, MA},\ \bibinfo {year} {1997})\ pp.\ \bibinfo {pages} {131--192}\BibitemShut {NoStop}%
\bibitem [{\citenamefont {Gubernatis}\ \emph {et~al.}(2016)\citenamefont {Gubernatis}, \citenamefont {Kawashima},\ and\ \citenamefont {Werner}}]{gubernatis_quantum_2016}%
  \BibitemOpen
  \bibfield  {author} {\bibinfo {author} {\bibfnamefont {J.}~\bibnamefont {Gubernatis}}, \bibinfo {author} {\bibfnamefont {N.}~\bibnamefont {Kawashima}},\ and\ \bibinfo {author} {\bibfnamefont {P.}~\bibnamefont {Werner}},\ }\href@noop {} {\emph {\bibinfo {title} {Quantum {{Monte Carlo Methods}}: {{Algorithms}} for {{Lattice Models}}}}}\ (\bibinfo  {publisher} {Cambridge University Press},\ \bibinfo {address} {Cambridge},\ \bibinfo {year} {2016})\BibitemShut {NoStop}%
\bibitem [{Note4()}]{Note4}%
  \BibitemOpen
  \bibinfo {note} {In the TEBD simulations, the Kraus-Cirac decomposition is not performed.}\BibitemShut {Stop}%
\bibitem [{\citenamefont {Fischer}\ \emph {et~al.}(2026)\citenamefont {Fischer}, \citenamefont {Leahy}, \citenamefont {Eddins}, \citenamefont {Keenan}, \citenamefont {Ferracin}, \citenamefont {Rossi}, \citenamefont {Kim}, \citenamefont {He}, \citenamefont {Pietracaprina}, \citenamefont {Sokolov}, \citenamefont {Dooley}, \citenamefont {Zimbor{\'a}s}, \citenamefont {Tacchino}, \citenamefont {Maniscalco}, \citenamefont {Goold}, \citenamefont {{Garc{\'i}a-P{\'e}rez}}, \citenamefont {Tavernelli}, \citenamefont {Kandala},\ and\ \citenamefont {Filippov}}]{fischer_dynamical_2026-1}%
  \BibitemOpen
  \bibfield  {author} {\bibinfo {author} {\bibfnamefont {L.~E.}\ \bibnamefont {Fischer}}, \bibinfo {author} {\bibfnamefont {M.}~\bibnamefont {Leahy}}, \bibinfo {author} {\bibfnamefont {A.}~\bibnamefont {Eddins}}, \bibinfo {author} {\bibfnamefont {N.}~\bibnamefont {Keenan}}, \bibinfo {author} {\bibfnamefont {D.}~\bibnamefont {Ferracin}}, \bibinfo {author} {\bibfnamefont {M.~A.~C.}\ \bibnamefont {Rossi}}, \bibinfo {author} {\bibfnamefont {Y.}~\bibnamefont {Kim}}, \bibinfo {author} {\bibfnamefont {A.}~\bibnamefont {He}}, \bibinfo {author} {\bibfnamefont {F.}~\bibnamefont {Pietracaprina}}, \bibinfo {author} {\bibfnamefont {B.}~\bibnamefont {Sokolov}}, \bibinfo {author} {\bibfnamefont {S.}~\bibnamefont {Dooley}}, \bibinfo {author} {\bibfnamefont {Z.}~\bibnamefont {Zimbor{\'a}s}}, \bibinfo {author} {\bibfnamefont {F.}~\bibnamefont {Tacchino}}, \bibinfo {author} {\bibfnamefont {S.}~\bibnamefont {Maniscalco}}, \bibinfo {author} {\bibfnamefont {J.}~\bibnamefont {Goold}}, \bibinfo {author} {\bibfnamefont {G.}~\bibnamefont {{Garc{\'i}a-P{\'e}rez}}}, \bibinfo {author} {\bibfnamefont {I.}~\bibnamefont {Tavernelli}}, \bibinfo {author} {\bibfnamefont {A.}~\bibnamefont {Kandala}},\ and\ \bibinfo {author} {\bibfnamefont {S.~N.}\ \bibnamefont {Filippov}},\ }\href {https://doi.org/10.1038/s41567-025-03144-9} {\bibfield  {journal} {\bibinfo  {journal} {Nat. Phys.}\ }\textbf {\bibinfo {volume} {22}},\ \bibinfo {pages} {302} (\bibinfo {year} {2026})}\BibitemShut {NoStop}%
\bibitem [{\citenamefont {Fishman}\ \emph {et~al.}(2022)\citenamefont {Fishman}, \citenamefont {White},\ and\ \citenamefont {Stoudenmire}}]{fishman_itensor_2022}%
  \BibitemOpen
  \bibfield  {author} {\bibinfo {author} {\bibfnamefont {M.}~\bibnamefont {Fishman}}, \bibinfo {author} {\bibfnamefont {S.}~\bibnamefont {White}},\ and\ \bibinfo {author} {\bibfnamefont {E.~M.}\ \bibnamefont {Stoudenmire}},\ }\href {https://doi.org/10.21468/SciPostPhysCodeb.4} {\bibfield  {journal} {\bibinfo  {journal} {SciPost Physics Codebases}\ ,\ \bibinfo {pages} {004}} (\bibinfo {year} {2022})}\BibitemShut {NoStop}%
\end{thebibliography}
%

\end{document}